\documentclass[a4paper,11pt]{article}
\usepackage{jheppub}
\usepackage[utf8]{inputenc}

\usepackage{lmodern}
\usepackage[T1]{fontenc}
\usepackage{lipsum}

\usepackage{tikz}
\usetikzlibrary{
decorations.markings
}
\tikzset{
  ->-/.style={
    decoration={
      markings,
      mark=at position #1 with {\arrow{latex}}},
    postaction={decorate}
  },
  ->-/.default=0.5
}
\tikzset{
    partial ellipse/.style args={#1:#2:#3}{
        insert path={+ (#1:#3) arc (#1:#2:#3)}
    }
}

\makeatletter
\if@titlepage
  \renewenvironment{abstract}{%
      \titlepage
      \null\vfil
      \@beginparpenalty\@lowpenalty
      \begin{center}%
        \bfseries \abstractname
        \@endparpenalty\@M
      \end{center}}%
      {\par\vfil\null\endtitlepage}
\else
  \renewenvironment{abstract}{%
      \if@twocolumn
        \section*{\abstractname}%
      \else
        \small
        \begin{center}%
          {\bfseries \abstractname\vspace{-.5em}\vspace{\z@}}%
        \end{center}%
        \quotation
      \fi}
      {\if@twocolumn\else\endquotation\fi}
\fi
\makeatother        

\usepackage{amsmath,amsfonts,amssymb,amsthm}
\usepackage{mathrsfs}
\usepackage{mathtools}
\usepackage{bm}
\usepackage{tensor}
\usepackage{slashed}
\newcommand{\mb}{\ensuremath{\mathbb}}
\newcommand{\mc}{\ensuremath{\mathcal}}
\newcommand{\mf}{\ensuremath{\mathfrak}}
\newcommand{\mh}{\ensuremath{\mathscr}}
\newcommand{\mr}{\ensuremath{\mathrm}}
\newcommand{\ms}{\ensuremath{\mathsf}}
\renewcommand{\v}[1]{\ensuremath{\bm{\mathbf{#1}}}}
\renewcommand{\d}{\mr{d}}

\newcommand{\f}{\frac}
\newcommand{\I}{\indices}
\DeclareMathOperator{\tr}{tr}
\DeclareMathOperator{\sign}{sign}

\title{Topological Correlators and Surface Defects from Equivariant Cohomology}

\begin{document}

\begingroup\parindent0pt
\begin{flushright}\footnotesize
\texttt{UUITP-17/20}\\
\end{flushright}
\vspace*{4em}
\centering
\begingroup\LARGE
\bf 
Topological Correlators and Surface Defects \\ from Equivariant Cohomology
\par\endgroup
\vspace{2.5em}
\begingroup\large
{\bf Rodolfo Panerai}, {\bf Antonio Pittelli}, {\bf Konstantina Polydorou}
\par\endgroup
\vspace{1em}
\begingroup\itshape
Department of Physics and Astronomy, Uppsala University, \\
Box 516, SE-75120 Uppsala, Sweden
\par\endgroup
\vspace{1em}
\vspace{2.5em}
\endgroup

\begin{abstract}
\noindent
We find a one-dimensional protected subsector of $\mathcal{N}=4$ matter theories on a general class of three-dimensional manifolds. By means of equivariant localization we identify a dual quantum mechanics  computing  BPS correlators of the original model in three dimensions. Specifically, applying the Atiyah--Bott--Berline--Vergne formula to the original action demonstrates  that this localizes on a one-dimensional action with support on the fixed-point submanifold of suitable isometries. We first show that our approach reproduces previous results obtained on $S^3$. Then, we apply it to the novel case of $S^2 \times S^1$ and show that the theory localizes on two noninteracting quantum mechanics with disjoint support. We prove that the BPS operators of such models are naturally associated with a noncommutative star product, while their correlation functions are essentially topological. Finally, we couple the three-dimensional theory to general $\mathcal{N}=(2,2)$ surface defects and extend the localization computation to capture the full partition function and BPS correlators of the mixed-dimensional system.
\end{abstract}


\thispagestyle{empty}

\newpage
\hrule
\tableofcontents
\afterTocSpace
\hrule
\afterTocRuleSpace

\section{Introduction and Summary}
Recent years have witnessed great progress in our understanding of the nonperturbative dynamics of supersymmetric field theories. A key factor has been the development of new computational techniques that lead to a large class of new exact results. Despite this progress, it has been proven hard to extend these techniques to compute correlators of local operators inserted at arbitrary points for theories in more than two dimensions. In the context of supersymmetric localization, for instance, this is typically not possible, as one is restricted to the cohomology induced by the localizing supercharge.

The situation is quite different in the case of superconformal theories, where, in addition to supersymmetry, one can make use of the constraining power of conformal invariance. An important result in this direction was presented in \cite{Beem:2013sza}, where, in the context of $\mc{N}=2$ SCFTs in four dimensions, a chiral-algebra structure, proper of two-dimensional models, was identified in the correlation functions of certain protected operators. These are so-called Schur operators, i.e.\ Higgs-branch operators belonging to the cohomology of a pair of suitable nilpotent supercharges mixing the R-symmetry. Similar constructions have unveiled vertex operator algebras hidden in six-dimensional $\mc{N} = (2,0)$ theories \cite{Beem:2014kka}, and one-dimensional topological sectors of $\mc{N} = 4$ SCFTs in three dimensions associated, respectively, with the Higgs and the Coulomb branch \cite{Chester:2014mea,Beem:2016cbd}.

This analysis, carried out at the level of the superconformal algebra, was later approached from a path-integral point of view. In the case of three dimensions, the authors of \cite{Dedushenko:2016jxl} were able, by means of supersymmetric localization, to obtain a Lagrangian description of the one-dimensional Higgs-branch sector for theories on $S^3$. Interestingly, the construction still holds when one introduces a certain mass deformation, thus breaking away from the realm of conformal theories. The same approach was later extended to the Coulomb-branch sector through mirror symmetry \cite{Dedushenko:2017avn,Dedushenko:2018icp}, and to five dimensions \cite{Mezei:2018url}. For other recent developments, see also \cite{Chang:2019dzt,Fan:2019jii,Gaiotto:2019mmf}.

Further results have been obtained in four dimensions. A duality between $\mc{N} = 2$ theories on $S^4$ and a symplectic boson system on $S^2$ was identified in \cite{Pan:2017zie}. More recently, a correspondence between a $\beta$-$\gamma$ system on the two-torus and $\mc{N} = 2$ theories on $S^3\times S^1$ was also established \cite{Pan:2019bor,Dedushenko:2019yiw}. The link between three and four dimensions was explored in \cite{Pan:2019shz,Dedushenko:2019mzv,Dedushenko:2019mnd}. Notably, for four-dimensional theories it has not yet been possible to deviate from the conformal point since the dualities explored in \cite{Pan:2017zie,Pan:2019bor,Dedushenko:2019yiw} make explicit use of features exclusive to superconformal field theories, such as non-trivial $\mr{U}(1)_{\mr{r}}$ R-symmetry background fields.

What these constructions have in common is the idea that protected sectors of the theories under consideration could be described by lower dimensional models with support on special submanifolds singled out by the cohomology induced by suitable nilpotent supercharges.
This suggests the possibility to address this problem by means of a geometric approach that could span diverse backgrounds. Indeed, this is one of the goals of the present paper. We apply this idea to $\mc{N}=4$ matter theories in three dimensions and derive a formula that generates the actions of the dual one-dimensional models. With this formula we study a geometric background that has not been previously considered in the literature, namely $S^2\times S^1$.
Partition functions of supersymmetric theories on $S^{d-1} \times S^1$ are referred to as $d$-dimensional supersymmetric indices. In particular, if the theory on $S^{d-1} \times S^1$ is conformal, the partition function provides its superconformal index. The fact that $d$-dimensional supersymmetric indices are encoded by $(d-2)$-dimensional models was first observed for $d=4$ by conveniently rewriting the corresponding matrix integrals \cite{Gadde:2011ik}. First-principles derivations based on cohomological arguments appeared later on by studying suitable limits of the superconformal index \cite{Beem:2013sza,Pan:2019bor}. A similar derivation for three-dimensional supersymmetric indices had not yet appeared in the literature: we fill such a gap in this paper by finding the quantum mechanics underlying a three-dimensional hypermultiplet index.
Moreover, it was found \cite{Chester:2014mea,Beem:2016cbd,Dedushenko:2016jxl} that $\mc{N}=4$ SCFTs in three dimensions enjoy a one-dimensional OPE algebra with an associated noncomutative star product (see also \cite{Etingof:2019guc} for a more mathematically-oriented perspective).
Indeed, we are able to identify such a structure in our quantum mechanics even in the presence of a mass deformation.

\paragraph{Summary of Results.} In the present work we consider a theory of $\mc{N}=4$ hypermultiplets conformally coupled to some three-manifold $\mh{M}$. The theory has generic R-symmetry twisting and weak gauging of the flavor current. If the algebra generated by a supercharge $\v{Q}$ contains a real isometry $\mc{L}_v$ we show that, when the action of the theory is restricted to $\v{Q}$-closed field configurations, it localizes on $S_{\mr{1d}}$, a one-dimensional action with support on $\mh{M}_v$, the fixed-point locus of $v$ in $\mh{M}$. Through the Atiyah--Bott--Berline--Vergne formula we are able to derive the general expression
\begin{align}\label{EQ:localization_introduction}
  S_{\mr{1d}} = \int_{\mh{M}_v} \f{2\tilde{\mf{q}}^{\dot{a}}(\epsilon_{\dot{a}\dot{c}}\mr{D}_\mu + \Phi_{\dot{a}\dot{c}}\gamma_\mu)\mf{q}^{\dot{c}}}{\mr{e}_v(N\mh{M}_v)}\;\mr{d}x^\mu \;,
\end{align}
where $\mathfrak q$ and $\tilde{\mathfrak q}$ are suitable contractions of Killing spinors on $\mh{M}$ with hypermultiplet scalars, $\mr{D}_\mu$ is a covariant derivative, $\Phi_{\dot{a}\dot{c}}$ are scalar fields of a background vector multiplet, and $\mr{e}_v(N\mh{M}_v)$ is the $v$-equivariant Euler class of the normal bundle $N\mh{M}_v$.
Along the lines of \cite{Dedushenko:2016jxl, Bonetti:2016nma}, we interpret the above as the action functional of a one-dimensional quantum theory that produces correlation functions of BPS operators inserted on $\mh{M}_v$.

The formula \eqref{EQ:localization_introduction} can be directly applied to various geometries. We first consider the case of $\mh{M} \simeq S^3$ and show that, by choosing some $v$ generating a circle action, we obtain a theory on $\mh{M}_v \simeq S^1$ that matches \cite{Dedushenko:2016jxl}. Thus, we turn our attention to the case of $\mh{M} \simeq S^2 \times S^1$, which is new. By selecting $v$ to be the azimuthal rotation on the two-sphere, we localize on a one-dimensional theory with support on two disjoint circles located at the north and south poles of $S^2$. Its action reads
\begin{align}\label{EQ:1d_action_index}
  S_{\mr{1d}} = 2\pi r \int_0^{2\pi} \tilde{q}^+(t)\,(\partial_t - \mr{i}\zeta)\,q^+(t) \;\mr{d}t + 2\pi r \int_0^{2\pi} \tilde{q}^-(t)\,(\partial_t - \mr{i}\zeta^*)\,q^-(t) \;\mr{d}t \;,
\end{align}
where the integration is over the two copies of $S^1$, the fields are protected Higgs-branch operators and $\zeta = a - \mr{i}\beta\sigma$ is some combination of fields in the background vector multiplet.

We study the path integral of such quantum mechanics by addressing the problem of identifying the correct integration cycle and by showing that its partition function precisely matches its three-dimensional counterpart.
Indeed, the quantum mechanics with action \eqref{EQ:1d_action_index} computes $Z_{\rm hyper}$, which is the supersymmetric index of a $\mathcal N=4$ hypermultiplet with a real mass deformation $\sigma$ and a flavor fugacity $a$. The hypermultiplet index given by $Z_{\rm hyper}$ is divergent whenever $\sigma = 0$ and $a\in\mb{Z}$. This is not inconsistent as supersymmetric indices are not supposed to be analytic functions of the fugacities, but rather meromorphic ones \cite{Rastelli:2016tbz}. More precisely, in four dimensions the residues of the index at poles in flavor fugacities were linked to indices of dual theories coupled to surface defects  \cite{Gaiotto:2012xa}.

We then use the theory to compute BPS correlators of Higgs-branch operators. Similarly to \cite{Dedushenko:2016jxl, Beem:2016cbd}, we find that these are essentially topological in nature and define a star product that induces a noncommutative algebra of operators.
Such a product acts on one-dimensional $\mf{su}(N)$ flavor symmetry currents $J^i{}_j$ with
\begin{align}
  J^i{}_j \ast J^k{}_l &= J^i{}_j\,J^k{}_l + \delta^i{}_l\,J^k{}_j - \delta^k{}_j\,J^i{}_l - \left(\delta^k{}_j\delta^i{}_l-\f{1}{N}\delta^i{}_j\delta^k{}_l\right) \;.
\end{align}
Finally, we introduce $\mc{N}=(2,2)$ defects with support on $S^2$ and couple them to the bulk theory through twisted superpotentials. Because $\v{Q}$ is compatible with the standard choice of localizing supercharge on the two-sphere, we are able to localize the resulting theory and write its partition function as that of a coupled system of zero- and one-dimensional quantum theories.

\paragraph{Outlook.} It would be interesting to extend the formalism developed in this paper to gauge theories with  dynamical vector multiplets. Especially, we would like to understand how BRST symmetry modifies \eqref{EQ:localization_introduction} and which one-dimensional models capture the degrees of freedom of three-dimensional vector multiplets. Incorporating gauge theories in our language would also allow to investigate the way mirror symmetry manifests itself at the level of one-dimensional actions.

Moreover, supersymmetric localization was successfully applied to theories defined on non-compact manifolds \cite{David:2016onq,David:2018pex,David:2019ocd,Pittelli:2018rpl} and on manifolds with boundaries \cite{Hori:2013ika,Yoshida:2014ssa,Gava:2016oep,Longhi:2019hdh}. Generalizing \eqref{EQ:localization_introduction} to such instances would be a natural direction to explore. In that case, it should be possible to make contact with  \cite{Bullimore:2016nji,Bullimore:2016hdc,Dimofte:2019zzj}, where the interplay between boundary conditions and cohomology classes of BPS operators was studied.

Furthermore, we expect that our construction has a direct uplift to higher dimensions. We wish to report on these points in future publications.

\paragraph{Outline on the Paper.}
In Section \ref{SEC:setup} we review three-dimensional $\mc{N}=4$ multiplets conformally coupled to curved backgrounds, their actions and the associated superalgebras.

In Section \ref{SEC:localization} we provide a detailed explanation of the cohomological approach we adopt to localize three-dimensional models down to one-dimensional ones. As an example, we apply our general formula to hypermultiplets on the three-sphere. The outcome is a one-dimensional theory on a great circle of $S^3$, in agreement with \cite{Dedushenko:2016jxl}.

In Section \ref{SEC:S2S1} we use our general formula upon a novel instance, namely hypermultiplets on $S^2 \times S^1$ with isometry superalgebra $\mathfrak{su}(2|2)$. We explicitly find the corresponding  Killing spinors as well as the background vector multiplet yielding a real mass deformation and a non-trivial flavor holonomy along $S^1$. In this case, the result of the cohomological localization is a one-dimensional theory on two circles sitting at antipodal points of the two-sphere.

In Section \ref{SEC:1d_theory} we use the one-dimensional action found in Section \ref{SEC:S2S1} to compute correlation functions of BPS operators built out of hypermultiplet scalars. Furthermore, we compute the partition function of the one-dimensional model and check that it matches the partition function of the original three-dimensional theory. Finally, we employ Morse theory, as an alternative to the BPS ansatz, to find an integration cycle for the path integral of the theory.

In Section \ref{SEC:correlators} we discuss the connection between the BPS correlators of the localized theory and those of an associated topological model. We introduce a noncommutative star product between operators and apply it in particular to the flavor Noether currents of the one-dimensional theory.

In Section \ref{SEC:defects} we couple the three-dimensional theory on $S^2 \times S^1$ to two-dimensional defects supported on $S^2$. Specifically, we find the subalgebra of $\mathfrak{su}(2|2)$ preserved by the defects and the corresponding two-dimensional Killing spinors. Then, we couple the bulk mulitplets to the defect multiplets through twisted superpotentials and proceed to localize the resulting theory.

\section{\texorpdfstring{$\mc{N}=4$ Supersymmetry in Three Dimensions}{N=4 Supersymmetry in Three Dimensions}}\label{SEC:setup}
We consider a three-dimensional theory with $N$ dynamical $\mc{N}=4$ hypermultiplets.
These have components
\begin{align}
  (q_a, \tilde{q}_a, \psi_{\dot{a}}, \tilde{\psi}_{\dot{a}}) \;.
\end{align}
For all fields, we keep explicit the indices associated with the R-symmetry algebra representation that acts on them. In particular, the scalars $q_a, \tilde{q}_a$ and the spinors $\psi_{\dot{a}}, \tilde{\psi}_{\dot{a}}$ belong, respectively, to the $(\v{2},\v{1})$ and $(\v{1},\v{2})$ representation of $\mf{su}(2)_{\mr{H}}\oplus\mf{su}(2)_{\mr{C}}$. Although flavor indices are kept implicit, one should keep in mind that the components of the $N$ hypermultiplets transform under the fundamental representation of the flavor group $\mr{USp}(2N)$. We will also refer to the flavor subgroup $\mr{U}(N)$, which is embedded in the fundamental representation of $\mr{USp}(2N)$ as $\v{N}\oplus\overline{\v{N}}$.

The multiplets are coupled to various background fields. The R-symmetry currents are coupled to background flat connections $A_\mr{H}$ and $A_\mr{C}$. The theory is then conformally coupled to the rigid geometry of a closed three-dimensional manifold $\mh{M}$ with metric $g$. Finally, the current associated with a subgroup $G$ of the $\mr{U}(N)$ flavor subgroup is coupled to a background vector multiplet whose top component is the $G$-connection $A$. We denote with $\mf{g}$ the Lie algebra of $G$ and with $\mc{R}:\mf{g}\to\mf{u}(N)$ its representation. All these background connections enter in the definition of the covariant derivative
\begin{align}
  \mr{D} = \nabla - \mr{i}A_{\mr{H}} - \mr{i}A_{\mr{C}} - \mr{i}A \;.
\end{align}

Superconformal transformations are generated by spinors $\xi^{a\dot{a}}$ that are solutions of the conformal Killing equation
\begin{align}\label{EQ:killing_spinor_equation}
  \mr{D}_\mu\xi^{a\dot{a}} = \gamma_\mu\eta^{a\dot{a}} \;,
\end{align}
for some spinor $\eta^{a\dot{a}}$. Both $\xi^{a\dot{a}}$ and $\eta^{a\dot{a}}$ belong to the $(\v{2},\v{2})$ representation of the R-symmetry algebra.
These define the supersymmetry transformations
\begin{align}
  \delta q^a &= \xi^{a\dot{a}}\psi_{\dot{a}} \;, \cr
  \delta \tilde{q}^a &= \xi^{a\dot{a}}\tilde{\psi}_{\dot{a}} \;, \label{EQ:susy_scalars}
\end{align}
and
\begin{align}
  \delta \psi_{\dot{a}} &= \mr{i}\gamma^\mu\xi_{a\dot{a}}\mr{D}_\mu q^a + \mr{i}\eta_{a\dot{a}}q^a - \mr{i}\xi_{a\dot{c}}{\Phi_{\dot{a}}}^{\dot{c}}q^a \;, \cr
  \delta \tilde{\psi}_{\dot{a}} &= \mr{i}\gamma^\mu\xi_{a\dot{a}}\mr{D}_\mu\tilde{q}^a + \mr{i}\tilde{q}^a\eta_{a\dot{a}} + \mr{i}\xi_{a\dot{c}}\tilde{q}^a{\Phi_{\dot{a}}}^{\dot{c}} \label{EQ:susy_fermions} \;,
\end{align}
which close on shell on the superconformal algebra $\mf{osp}(4|4)$ or a subalgebra thereof, according to $\mh{M}$.

The theory has action
\begin{align}\label{EQ:on_shell_action}
  S_{\mr{on}} = \int_{\mh{M}}\star\mh{L}_{\mr{on}}
\end{align}
where
\begin{align}
  \mh{L}_{\mr{on}} =\; &\mr{D}^{\mu}\tilde{q}^a\mr{D}_{\mu}q_a - \mr{i}\tilde{\psi}^{\dot{a}}\gamma^{\mu}\mr{D}_{\mu}\psi_{\dot{a}} + R/8\,\tilde{q}^aq_a 
  - \tilde{q}^a(\mr{i}D_{ac} + 1/2\,\epsilon_{ac}\Phi^{\dot{a}\dot{c}}\Phi_{\dot{a}\dot{c}})q^c + \mr{i}\tilde{\psi}_{\dot{a}}\Phi^{\dot{a}\dot{c}}\psi_{\dot{c}} \;.
\end{align}
The scalar fields $D_{ac}$ and $\Phi_{\dot{a}\dot{c}}$ are, together with the connection $A$, the bosonic degrees of freedom of the background vector multiplet. These take value in the Lie algebra $\mf{g}$ of $G$ and belong, respectively, to the $(\v{3},\v{1})$, $(\v{1},\v{3})$ and $(\v{1},\v{1})$ representations of the R-symmetry algebra. The mass term for the scalars, proportional to the Ricci curvature $R$, comes from the conformal coupling with the geometry.

The action above is invariant under \eqref{EQ:susy_scalars} and \eqref{EQ:susy_fermions}, provided that a BPS condition is imposed on the background fields. In fact, the variation of the action, according to the transformations above, reads
\begin{align}
  \delta S_{\mr{on}} = \int_{\mh{M}} \star(\mr{i}\tilde{q}^a\delta\lambda_{a\dot{a}}\psi^{\dot{a}} - \mr{i}\tilde{\psi}^{\dot{a}}\delta\lambda_{a\dot{a}}q^a)
\end{align}
where
\begin{align}
  \delta\lambda_{a\dot{a}} = -\mr{i}/2\,\epsilon^{\mu\nu\rho}\gamma_\rho\xi_{a\dot{a}}F_{\mu\nu} - {D_a}^c\xi_{c\dot{a}} - \mr{i}\gamma^\mu{\xi_a}^{\dot{c}}\mr{D}_\mu\Phi_{\dot{c}\dot{a}} + 2\mr{i}{\Phi_{\dot{a}}}^{\dot{c}}\eta_{a\dot{c}} + \mr{i}/2\,\xi_{a\dot{d}}[\Phi\I{_{\dot{a}}^{\dot{c}}},\Phi\I{_{\dot{c}}^{\dot{d}}}] \;,
\end{align}
is the expression of the supersymmetric variation of the gaugini for a dynamical vector multiplet.
The action is therefore supersymmetric for any choice of background fields that satisfy
\begin{align}\label{EQ:vector_bps}
  \delta \lambda_{a\dot{a}} = 0 \;.
\end{align}
Furthermore, we impose the following reality conditions on the background fields
\begin{align}\label{EQ:vector_reality_conditions}
  (A^B_\mu)^* &= +A^B_\mu \;, \cr
  (\Phi^B_{\dot{a}\dot{c}})^* &= -\Phi^{\dot{a}\dot{c}\,B} \;, \cr
  (D^B_{ac})^* &= -D^{ac\,B} \;,
\end{align}
where we have indicated with $B$ the Lie-algebra index associated with a given basis $\{\mr{t}_B\}$ for $\mf{g}$.

The supersymmetry algebra generated by the transformations \eqref{EQ:susy_scalars} and \eqref{EQ:susy_fermions} closes on shell. The reader can find a detailed account of this in Appendix \ref{APP:algebra_closure}. No known finite set of auxiliary fields can bring off shell the entire representation of the superconformal algebra acting on the multiplet. However, it is possible to bring off shell subsets of the algebra, and this will be a crucial step in our localization procedure. This can be realized by considering a modified Lagrangian,
\begin{align}\label{EQ:on_shell_lagrangian}
  \mh{L} = \mh{L}_{\mr{on}} + \tilde{G}^aG_a \;,
\end{align}
where two auxiliary fields, $G_a$ and $\tilde{G}_a$ have been introduced. These belong to the same representations of the scalars $q_a$ and $\tilde{q}_a$, respectively. Supersymmetry acts on them with
\begin{align}
  \delta G^a &= \nu^{a\dot{a}}(\gamma^\mu\mr{D}_\mu\psi_{\dot{a}} - \Phi_{\dot{a}\dot{c}}\psi^{\dot{c}}) \;, \cr
  \delta \tilde{G}^a &= \nu^{a\dot{a}}(\gamma^\mu\mr{D}_\mu\tilde{\psi}_{\dot{a}} + \tilde{\psi}^{\dot{c}}\Phi_{\dot{a}\dot{c}}) \;,
\end{align}
and to accommodate for their introduction, we also need to modify the variation of the fermions as
\begin{align}\label{EQ:susy_fermions_off_shell}
  \delta \psi_{\dot{a}} &= \mr{i}\gamma^\mu\xi_{a\dot{a}}\mr{D}_\mu q^a + \mr{i}\eta_{a\dot{a}}q^a - \mr{i}\xi_{a\dot{c}}{\Phi_{\dot{a}}}^{\dot{c}}q^a - \mr{i}\nu_{a\dot{a}}G^a \;, \cr
  \delta \tilde{\psi}_{\dot{a}} &= \mr{i}\gamma^\mu\xi_{a\dot{a}}\mr{D}_\mu\tilde{q}^a + \mr{i}\tilde{q}^a\eta_{a\dot{a}} + \mr{i}\xi_{a\dot{c}}\tilde{q}^a{\Phi_{\dot{a}}}^{\dot{c}} - \mr{i}\nu_{a\dot{a}}\tilde{G}^a \;.
\end{align}
The above are written in terms of the auxiliary spinors $\nu^{a\dot{a}}$, which should be intended as functions of $\xi^{a\dot{a}}$.
To ensure off-shell closure, the auxiliary spinors should satisfy
\begin{align}\label{EQ:auxiliary_identities}
  {\xi_a}^{\dot{u}}\nu_{b\dot{u}} &= 0 \;, \cr
  ({\nu^c}\vphantom{\nu}_{\dot{a}})_{\alpha}\,(\nu_{c\dot{c}})_{\beta} &= ({\xi^c}\vphantom{\nu}_{\dot{c}})_{\alpha}\,(\xi_{c\dot{a}})_{\beta} \;, \cr
  {\nu_{(a}}^{\dot{u}}\gamma^\mu\mr{D}_\mu\nu_{b)\dot{u}} &= -2\mr{i}/3\,{\xi_{(a}}^{\dot{u}}\gamma^\mu\mr{D}_\mu\xi_{b)\dot{u}} \;.
\end{align}

To quantize the theory, we specify reality conditions for the bosonic degrees of freedom in the path integral,
\begin{align}\label{EQ:hyper_reality_conditions}
  \tilde{q}^a &= (q_a)^* \;, \cr
  \tilde{G}^a &= (G_a)^* \;.
\end{align}
The conditions above guarantee that the bosonic part of the action is positive semidefinite.

\section{Cohomological Approach}\label{SEC:localization}
As explained in Appendix \ref{APP:algebra_closure}, a given supercharge $\mathbf{Q}$ associated with a conformal Killing spinor $\xi^{a\dot{a}}$ squares to a combination of various bosonic symmetries. Among these, there is a conformal isometry generated by the conformal Killing vector $v$
\begin{align}\label{EQ:conformal_killing_vector}
  v^\mu = \mr{i}\xi^{a\dot{a}}\gamma^\mu\xi_{a\dot{a}} \;.
\end{align}
In our analysis, we will focus on a certain subset of supercharges. In particular, we choose a $\xi^{a\dot{a}}$ such that the associated $v$ is real and has vanishing divergence,
\begin{align}
  0 = \operatorname{div}v = 6\mr{i}\xi^{a\dot{a}}\eta_{a\dot{a}} \;.
\end{align}
Such a $v$ obeys the Killing equation
\begin{align}
  \mc{L}_v g = 0
\end{align}
and generates proper real isometries on $\mh{M}$.

We also introduce two scalars, defined with
\begin{align}
  X_{ac} &= {\xi_{a}}^{\dot{a}}\xi_{c\dot{a}} \;, & \bar{X}_{\dot{a}\dot{c}} &= \xi^{a}\vphantom{\xi}_{\dot{a}}\xi_{a\dot{c}} \;,
\end{align}
which respectively belong to the $(\v{3},\v{1})$ and the $(\v{1},\v{3})$ representation of $\mf{su}(2)_{\mr{H}}\oplus\mf{su}(2)_{\mr{C}}$. These obey the identities
\begin{align}\label{EQ:determinant_formula}
  4X_{au}X^{uc} &= {\delta_a}^c|v|^2 \;, & 4\bar{X}_{\dot{a}\dot{u}}\bar{X}^{\dot{u}\dot{c}} &= {\delta_{\dot{a}}}^{\dot{c}}|v|^2 \;.
\end{align}

Let us then consider the set of points where $v$ vanishes, i.e.\ the set of fixed points of the circle-group action generated by $v$. This is a submanifold $\mh{M}_v$ that can be decomposed as the union
\begin{align}
  \mh{M}_v = \bigcup_i \mh{M}_i \;,
\end{align}
in terms of disjoint $\mh{M}_i\simeq S^1$ \cite{Kobayashi:1972}.
Our goal is to show that, when we impose the BPS condition
\begin{align}\label{EQ:hyper_bps}
  \psi_{\dot{a}} &= 0 \;, &  \delta\psi_{\dot{a}} &= 0 \;, \cr
  \tilde{\psi}_{\dot{a}} &= 0 \;, &  \delta\tilde{\psi}_{\dot{a}} &= 0 \;,
\end{align}
the action
\begin{align}\label{EQ:off_shell_action}
  S = \int_{\mh{M}}\star\mh{L}
\end{align}
localizes on an action for a one-dimensional scalar theory with support on $\mh{M}_v$.
This can be shown in the context of (abelian) equivariant cohomology. We introduce the $v$-equivariant differential
\begin{align}
  \mr{d}_v = \mr{d} - \iota_v
\end{align}
and
\begin{align}
  \Omega = \star\mh{L} + \alpha_1 \;,
\end{align}
the sum of the 3-form Lagrangian $\star\mh{L}$ and the one-form $\alpha_1$ defined as
\begin{align}\label{EQ:one_form_action}
  \alpha_1 = X_{ac}(\tilde{q}^a\mr{D}q^c - \mr{D}\tilde{q}^aq^c) + \tilde{q}^aq_a\,w + 2\tilde{q}^a\Lambda_{ac}q^c
\end{align}
where
\begin{align}
  w_\mu &= \xi^{u\dot{u}}\gamma_{\mu}\eta_{u\dot{u}} \;, \cr
  (\Lambda_{ac})_\mu &= (\xi_{a\dot{a}}\gamma_{\mu}\xi_{c\dot{c}})\Phi^{\dot{a}\dot{c}} \;.
\end{align}
One can show that, on the solutions of the BPS equations \eqref{EQ:vector_bps}, the polyform $\Omega$ is equivariantly closed, i.e.
\begin{align}
  \mr{d}_v\Omega = 0 \;.  
\end{align}
This is described in more detail in Appendix \ref{APP:1d_action}.

This crucial observation implies that, on those solutions, the action \eqref{EQ:off_shell_action} receives contributions only from the fixed-point submanifold $\mh{M}_v$. Moreover, one can explicitly compute the action on the BPS solutions by means of the Atiyah--Bott--Berline--Vergne formula \cite{Pestun:2016jze,berline1982classes,Atiyah:1984px} as an integral over $\mh{M}_v$,
\begin{align}\label{EQ:localization_formula}
  S|_{\mr{BPS}} = \int_{\mh{M}_v} \f{i^*\alpha_1}{\mr{e}_v(N\mh{M}_v)} \;.
\end{align}
Here $i\colon\mh{M}_v\hookrightarrow\mh{M}$ is the immersion of the fixed points locus in $\mh{M}$ and $\mr{e}_v(N\mh{M})$ is the $v$-equivariant Euler class of the normal bundle $N\mh{M}_v$.

Interestingly, the BPS action can be rewritten in a suggestive form by introducing fields
\begin{align}
  \mf{q}_{\dot{a}} &= \xi_{a\dot{a}}q^a \;, \cr
  \tilde{\mf{q}}_{\dot{a}} &= \xi_{a\dot{a}}\tilde{q}^a \;.
\end{align}
In terms of the above, the one-form Lagrangian \eqref{EQ:one_form_action} reads
\begin{align}
  \alpha_1 = 2\tilde{\mf{q}}^{\dot{a}}\mf{D}_{\dot{a}\dot{c}}\mf{q}^{\dot{c}} - \mr{d}(\tilde{\mf{q}}^{\dot{a}}\mf{q}_{\dot{a}}) \;,
\end{align}
where
\begin{align}
  (\mf{D}_{\dot{a}\dot{c}})_\mu = \epsilon_{\dot{a}\dot{c}}\mr{D}_\mu + \Phi_{\dot{a}\dot{c}}\gamma_\mu
\end{align}
and the exact term can be dropped when integrating over $\mh{M}_v$, which is compact.

With the identity \eqref{EQ:localization_formula} established, it is tempting to interpret the integral in \eqref{EQ:localization_formula} as the action of a quantum one-dimensional theory that captures the correlators of BPS operators in the original theory in three dimensions. A similar approach has been taken in \cite{Dedushenko:2016jxl, Bonetti:2016nma, Pan:2017zie, Mezei:2018url, Costello:2018txb, Pan:2019bor}.

In the conventional approach to supersymmetric localization, one starts with a $\v{Q}$-exact deformation
\begin{align}
  S(t) = S_0 + t\v{Q}V
\end{align}
of the original action $S_0$, chosen in such a way that its bosonic part, when the appropriate reality conditions are taken into account, has positive semidefinite real part. This condition is necessary to guarantee the convergence of the path integral, which can be then represented as an integral restricted over the sole space of BPS solutions. In this integral, the original action evaluated on the BPS locus appears together with a term that captures the fluctuations of the dynamical fields around the BPS solutions, at the first order in the $t^{-1}$ expansion. While the action term depends only on the choice of the supercharge $\mathbf{Q}$,\footnote{
  While the field configurations that minimize $V$ can vary with the deformation term itself, these must form an improper subset of those configurations that are $\v{Q}$-closed. In this sense, formula \eqref{EQ:localization_formula} is universal.
} the one-loop determinant is related to the particular deformation term $V$ considered.

In the context of this localization scheme, however, the computation of the latter turns out to be a nontrivial task. At the same time, by direct comparison with known results, the one-loop determinant has been shown in previous cases \cite{Dedushenko:2016jxl, Pan:2017zie, Mezei:2018url, Pan:2019bor} to bring a trivial contribution to the overall computation.

Our approach here is close in spirit to the one of \cite{Bonetti:2016nma}. We argue that the result in \eqref{EQ:localization_formula} is more general, since it does not rely on the choice of a localizing action $\v{Q}V$. In fact, it does not even rely on the particular choice of reality conditions imposed on the fields.\footnote{The identities in \eqref{EQ:hyper_reality_conditions} and \eqref{EQ:vector_reality_conditions} will be used throughout the rest of the paper, but do not enter the proof in Appendix \ref{APP:1d_action}.} We refrain from giving a general prescription on how to determine the one-loop contribution to the partition function.

\subsection{\texorpdfstring{An Example: Hypermultiplets on $S^3$}{An Example: Hypermultiplets on S3}}
To better illustrate the localization prescription, we start by considering a known case originally studied in \cite{Dedushenko:2016jxl}. We will show how known results can be recovered through the formalism introduced above.

We consider the case where $\mh{M}\simeq S^3$. The metric
\begin{align}
  g = \delta_{\ms{a}\ms{b}} \; e^{\ms{a}} \otimes e^{\ms{b}} \;,
\end{align}
is determined by the choice of dreibein
\begin{align}
  e^{\ms{1}} &= r\sin(\theta) \, \mr{d}\varphi \;, \cr
  e^{\ms{2}} &= r\cos(\theta) \, \mr{d}\tau \;, \cr
  e^{\ms{3}} &= r \, \mr{d}\theta \;,
\end{align}
and coordinates $\theta\in[0,\pi/2]$, $\tau\in[-\pi,\pi)$, $\varphi\in[-\pi,\pi)$. Both background R-symmetry connections $A_{\mr{H}}$ and $A_{\mr{C}}$ are taken to be vanishing. Following \cite{Dedushenko:2016jxl}, we consider the gauge background
\begin{align}
\Phi_{\dot{1}\dot{2}} &= r^{-1}\sigma \;, \cr
D_{11} = D_{22} &= -\mr{i}r^{-2}\sigma \;.
\end{align}

The localizing supercharge that we are going to adopt is generated by a family of Killing spinors $\xi^{a\dot{a}}$, parametrized by $\beta$, with components
\begin{align}
  \xi^{1\dot{1}} &=  e^{-\f{\mr{i}}{2}(\theta-\tau+\varphi)}\begin{pmatrix}
    -1+\mr{i}e^{\mr{i}(\theta+\varphi)}\\
    \mr{i} e^{\mr{i}\theta} - e^{\mr{i}\varphi}
  \end{pmatrix} \;, \cr
  \xi^{1\dot{2}} &= -\f{\beta}{8} e^{-\f{\mr{i}}{2}(\theta+\tau+\varphi)}\begin{pmatrix}
    e^{\mr{i}\theta} +\mr{i} e^{\mr{i}\varphi}\\
    -\mr{i}-e^{\mr{i}(\theta+\varphi)}
  \end{pmatrix} \;, \cr
  \xi^{2\dot{1}} &=  e^{-\f{\mr{i}}{2}(\theta-\tau+\varphi)}\begin{pmatrix}
   -\mr{i}+e^{\mr{i}(\theta+\varphi)}\\
  -e^{\mr{i}\theta} +\mr{i} e^{\mr{i}\varphi}
  \end{pmatrix} \;, \cr
  \xi^{2\dot{2}} &= -\f{\beta}{8} e^{-\f{\mr{i}}{2}(\theta+\tau+\varphi)}\begin{pmatrix}
     \mr{i}e^{\mr{i}\theta} + e^{\mr{i}\varphi}\\
    1+\mr{i}e^{\mr{i}(\theta+\varphi)}
  \end{pmatrix} \;.
\end{align}
These have auxiliary spinors $\nu^{a\dot{a}}$ defined with
\begin{align}
\nu^{a\dot{a}} =\xi^{a\dot{a}}|_{\beta\mapsto-\beta} \;.
\end{align}
The associated isometry is generated by the vector field
\begin{align}
  v = 2r^{-1}\beta\,\partial_\tau \;,
\end{align}
and its square modulus
\begin{align}
  |v|^2 = 4\beta^2\cos^2\!\theta
\end{align}
vanishes on the circle $\mh{M}_v\simeq S^1$ that sits at $\theta = \pi/2$.

Furthermore, we have
\begin{align}
  X_{ac} &= \beta
  \begin{pmatrix}
    -1+\sin\theta\cos\varphi & \sin\theta\sin\varphi \\
    \sin\theta\sin\varphi & -1-\sin\theta\cos\varphi
  \end{pmatrix} \;, \cr
  \bar{X}_{\dot{a}\dot{c}} &= \mr{i}\cos\theta
  \begin{pmatrix}
    \f{1}{8}\beta^2 e^{-\mr{i}\tau} & 0 \\
    0 & -8 e^{\mr{i}\tau}
  \end{pmatrix} \;,
\end{align}
and
\begin{align}
  w &= \beta\sin^2\!\theta\,\mr{d}\varphi \;, \cr
  \Lambda_{ac} &=
  -\beta\sigma \sin\theta
  \begin{pmatrix}
    \sin\theta-\cos\varphi& -\sin\varphi \\
   - \sin\varphi & \sin\theta+\cos\varphi
  \end{pmatrix} \mr{d}\varphi
  -\beta\sigma \cos\theta
  \begin{pmatrix}
    -\sin\varphi & \cos\varphi \\
    \cos\varphi & \sin\varphi
  \end{pmatrix} \mr{d}\theta \;. \cr
\end{align}
As we will explain at the end of the next section, we construct the BPS operators from the null eigenvectors of $X_{ac}$, when evaluated on $\mh{M}_v$. These give
\begin{align}
  Q &= \cos(\varphi/2)\,q_1+ \sin(\varphi/2)\,q_2 \;, \cr 
  \tilde{Q} &= \cos(\varphi/2)\,\tilde{q}_1 + \sin(\varphi/2)\,\tilde{q}_2 \;.
\end{align}
The one-dimensional action is obtained from the embedding
\begin{align}
  i^*\alpha_1 = -2\beta\left[\tilde{Q}(\partial_t+\sigma)Q - (\partial_t-\sigma)\tilde{Q}Q\right] \mr{d}\varphi \;.
\end{align}
The geometric factor coming from $\mr{e}_v(N\mh{M}_v)$ brings a constant term. It is then immediate to show that the action, restricted to the BPS solutions, reads
\begin{align}
  S|_{\mr{BPS}} = -4\pi r\int_{-\pi}^{\pi}\mr{d}\varphi \; \tilde{Q}(\partial_\varphi+\sigma)Q \;,
\end{align}
which is the result found in \cite{Dedushenko:2016jxl}.

\section{\texorpdfstring{The Case of $S^2 \times S^1$}{The Case of S2xS1}}\label{SEC:S2S1}
\subsection{Supersymmetric Background}\label{SEC:S2S1_background}
\begin{figure}[ht]
  \centering
  \begin{tikzpicture}[thick,scale=1.3,>=latex]
  \begin{scope}[xshift=-2.1cm]
    \path[use as bounding box] (-1.6,-1.6) rectangle (1.6,1.6);
    \colorlet{shadow}{black!10}
    \begin{scope}
      \clip (-1.5,-1.5) rectangle (1.5,1.5);
      \fill [shadow] (0,0) circle[radius=1.5];
      \fill [white] (-0.1,0.5) circle[radius=1.7];
    \end{scope}
    \draw[->-=0.53, black!70] (0,0) [partial ellipse=-180:0:1.5 and 0.68];
    \draw[dashed, black!70] (0,0) [partial ellipse=0:+180:1.5 and 0.68];
    \draw[->-=0.25, black!70, rotate=115] (0,0) [partial ellipse=-35:-180:1.5 and 1.0];
    \draw[dashed, black!70, rotate=115] (0,0) [partial ellipse=145:+180:1.5 and 1.0];
    \draw[ultra thick, black] (0,0) circle (1.5);
    \fill [black](0,1.35) ellipse (1.5pt and 1pt);
    \fill [black!50](0,-1.35) ellipse (1.5pt and 1pt);
    \node at (0.8,1) {$\theta$};
    \node at (0,-0.4) {$\varphi$};
  \end{scope}
    \node at (0,0) {$\times$};
    \draw[ultra thick, black] (1,0) circle (0.5);
    \draw[->, black!70] (1,0) [partial ellipse=-120:-60:0.7 and 0.7];
    \node at (1,-1) {$t$};
  \end{tikzpicture}
  \caption{The choice of coordinates on $\mh{M}$.}\label{FIG:s2s1_coordinates}
\end{figure}
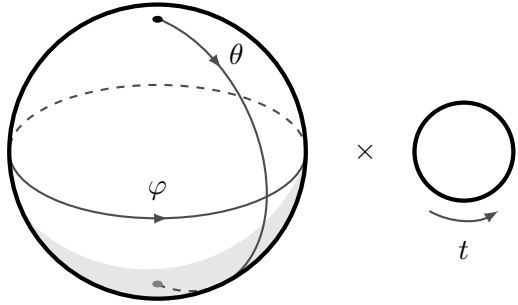

We now consider the case of $\mh{M}\simeq S^2\times S^1$, where the metric
\begin{align}
  g = \delta_{\ms{a}\ms{b}} \; e^{\ms{a}} \otimes e^{\ms{b}}
\end{align}
is expressed in terms of the dreibein
\begin{align}
  e^{\ms{1}} &= r\sin\theta\,\mr{d}\varphi\;, \cr
  e^{\ms{2}} &= r\,\mr{d}\theta \;, \cr
  e^{\ms{3}} &= r\beta\,\mr{d}t \;.
\end{align}
We have adopted coordinates $\theta\in[0,\pi]$, $\varphi\in[0,2\pi)$ and $t\in[0,2\pi)$, as in Figure \ref{FIG:s2s1_coordinates}.

The background R-symmetry connections are given by
\begin{align}
  (A_{\mr{H}})\I{^a_c} &= -\f{\mr{i}\beta}{2}(\sigma^3)\I{^a_c}\,\mr{d}t \;, \\
  (A_{\mr{C}})\I{_{\dot{a}}^{\dot{c}}} &= 0 \;.
\end{align}
In Appendix \ref{APP:background_gauge}, we derive a generic expression for the field components of the background vector multiplet satisfying the BPS condition \eqref{EQ:vector_bps}.
Here, we summarize the results by mentioning the nonvanishing components,
\begin{align}\label{EQ:vector_background_sols}
  A &= a\,\mr{d}t \;, \cr
  \Phi_{\dot{1}\dot{2}} &= r^{-1}\sigma \;, \cr
  D_{11} = D_{22} &= -\mr{i}r^{-2}\sigma \;,
\end{align}
written in terms of constants $a$ and $\sigma$ valued in $\mf{g}$. These combine into the complex combination
\begin{align}\label{EQ:zeta_definition}
  \zeta = a-\mr{i}\beta\sigma  
\end{align}
that will play an important role in the following.

The reader might be more familiar with a different set of BPS configurations that include, in addition to flat connections $a$, monopole solutions on the two-sphere labelled by monopole charges $\mf{m}$. These are the configurations that one finds when performing Coulomb-branch localization \cite{Imamura:2011su}, where the formula for the superconformal index is given in terms of an integral over $a$ and a sum over $\mf{m}$. The discrepancy between this locus and the one in \eqref{EQ:vector_background_sols} is a simple consequence of the fact that different localizing supercharges are employed in the localization procedure.
From the point of view of the weakly-gauged theory, this means that the hypermultitplet partiton function we consider will carry a dependence on the mass deformation $\sigma$, rather than on the monopole charges $\mf{m}$. Conversely, one should find monopole backgrounds when considering theories with twisted multiplets.
With the idea of extending our construction to theories with dynamical vector multiplets, the solutions in \eqref{EQ:vector_background_sols} seem to hint at a novel representation of the full supersymmetric index, where the end result is given in terms of an integral over $\zeta$ and $\zeta^*$. We leave this analysis for future work.

We consider the following set of Killing spinors
\begin{align}\label{EQ:killing_solutions}
  \xi^{1\dot{a}} &= +\gamma^{\ms{3}}e^{\f{\mr{i}}{2}\theta\gamma^{\ms{1}}}e^{-\f{\mr{i}}{2}\varphi\gamma^{\ms{3}}}\xi^{1\dot{a}}_0 \;, \cr
  \xi^{2\dot{a}} &= -e^{\f{\mr{i}}{2}\theta\gamma^{\ms{1}}}e^{-\f{\mr{i}}{2}\varphi\gamma^{\ms{3}}}\gamma^{\ms{3}}\xi^{2\dot{a}}_0 \;,
\end{align}
spanned by constant spinors $\xi_0^{a\dot{a}}$. These are particular solutions of the Killing spinor equation \eqref{EQ:conformal_killing_vector} that satisfy
\begin{align}\label{EQ:killing_condition}
  \eta^{a\dot{a}} = -\mr{i}(A_{\mr{H},\mu})^a\vphantom{)}_c\gamma^{\mu}\xi^{c\dot{a}} \;.
\end{align}

\subsection{Superalgebra}
The Killing spinors solutions in \eqref{EQ:killing_solutions}, parametrized be the 8 constants $(\xi_0^{a\dot{a}})_\alpha$, generate a special subalgebra of the superconformal algebra $\mf{osp}(4|4)$. This can be seen from the identity \eqref{EQ:killing_condition}; which, in the language of Appendix \ref{APP:algebra_closure}, implies
\begin{align}
  \operatorname{div} v &= 0 \;, \cr
  \rho &= 0 \;, \cr
  R^a{}_c-\mr{i}\iota_v(A_{\mr{H}})^a{}_c &= 0 \;.
\end{align}
This means that the resulting superalgebra, $\mf{su}(2|2)$, does not contain generators of conformal or $\mf{su}(2)_{\mr{H}}$ transformations. Its maximal bosonic subalgebra, $\mf{su}(2)\oplus\mf{u}(1)\oplus\mf{su}(2)_{\mr{C}}$ is the sum of the isometry algebra $\mf{su}(2)\oplus\mf{u}(1)$, generated by Killing vectors
\begin{align} 
  j_{\pm} &= - e^{\pm\mr{i}\varphi} (\pm\partial_\theta + \mr{i}\cot\theta\,\partial_\varphi) \;, \cr
  j_3 &= -\mr{i}\partial_\varphi \;, \cr
  z &= -\mr{i}\beta^{-1}\partial_t \;,
\end{align}
with
\begin{align}
  [j_3,j_\pm] &= \pm j_\pm \;, \cr
  [j_+,j_-] &= 2j_3 \;, \cr
  [z, j_\bullet] &= 0 \;,
\end{align}
and the $\mf{su}(2)_{\mr{C}}$ R-symmetry algebra
\begin{align}
  [\ms{R}_{\dot{a}\dot{b}}, \ms{R}_{\dot{c}\dot{d}}] = \epsilon_{\dot{a}\dot{d}}\ms{R}_{\dot{c}\dot{b}} - \epsilon_{\dot{c}\dot{b}}\ms{R}_{\dot{a}\dot{d}} \;.
\end{align}

Let us consider the action of supersymmetry on gauge-invariant operators. We adopt the following conventions for the supercharges. First, we define supercharges $\mc{Q}_{a\dot{a}}$ from the variations generated by the solutions in \eqref{EQ:killing_solutions} as
\begin{align}
  \delta\mc{O} &= \mr{i}\xi_0^{a\dot{a}}\mc{Q}_{a\dot{a}}\mc{O} \;.
\end{align}
Then, we define
\begin{align}
  \ms{Q}_{\dot{a}} &= (\mc{Q}_{1\dot{a}})_+ \;, & \ms{S}_{\dot{a}} &= (\mc{Q}_{2\dot{a}})_- \;, \cr
  \tilde{\ms{Q}}_{\dot{a}} &= (\mc{Q}_{2\dot{a}})_+ \;, & \tilde{\ms{S}}_{\dot{a}} &= (\mc{Q}_{1\dot{a}})_- \;.
\end{align}
These assignments ensure that $\ms{Q}_{\dot{a}}$ and $\tilde{\ms{Q}}_{\dot{a}}$ anticommute to positive roots of the isometry algebra, while $\ms{S}_{\dot{a}}$ and $\tilde{\ms{S}}_{\dot{a}}$ anticommute to negative roots. Specifically,
\begin{align}
  \{\ms{Q}_{\dot{a}}, \tilde{\ms{Q}}_{\dot{c}}\} &= -r^{-1}\epsilon_{\dot{a}\dot{c}}\,\ms{J}_+ \;, \cr
  \{\ms{S}_{\dot{a}}, \tilde{\ms{S}}_{\dot{c}}\} &= -r^{-1}\epsilon_{\dot{a}\dot{c}}\,\ms{J}_- \;, \cr
  \{\ms{Q}_{\dot{a}}, \ms{S}_{\dot{c}}\} &= -r^{-1}\epsilon_{\dot{a}\dot{c}}[\mr{i}\ms{J}_3+\ms{Z}] - \mr{i}r^{-1} \ms{R}_{\dot{a}\dot{c}} \;, \cr
  \{\tilde{\ms{Q}}_{\dot{a}}, \tilde{\ms{S}}_{\dot{c}}\} &= -r^{-1}\epsilon_{\dot{a}\dot{c}}[\mr{i}\ms{J}_3-\ms{Z}] - \mr{i}r^{-1}\ms{R}_{\dot{a}\dot{c}} \;,
\end{align}
where we have defined
\begin{align}
  \ms{J}_\bullet &= -\mc{L}_{j_\bullet} \;, & \ms{Z} &= -\mc{L}_{z} \;,
\end{align}
and where $\ms{R}_{\dot{a}\dot{b}}$ act on $\mf{su}(2)_{\mr{C}}$ indices as $(\ms{R}_{\dot{a}\dot{b}})_{\dot{c}\vphantom{\dot{b}}}{}^{\dot{d}}$, with
\begin{align}
  \ms{R}_{\dot{1}\dot{1}} &= \f{-\sigma^1+\mr{i}\sigma^2}{2} \;, &
  \ms{R}_{\dot{2}\dot{2}} &= \f{+\sigma^1+\mr{i}\sigma^2}{2} \;, &
  \ms{R}_{\dot{1}\dot{2}} = \ms{R}_{\dot{2}\dot{1}} &= \f{\sigma^3}{2} \;.
\end{align}

\subsection{Geometric Localization}\label{SEC:S2S1_localization}
For the purpose of localizing the theory, we focus on a $\widehat{\mf{su}}(1|1)$ subalgebra of the $\mf{su}(2|2)$ algebra considered above. This is generated by the supercharges
\begin{align}\label{EQ:su(1|1)_generators}
  \ms{Q} &= \ms{Q}_{\dot{1}} + \tilde{\ms{Q}}_{\dot{1}} \;, \cr
  \ms{S} &= \ms{S}_{\dot{2}} + \tilde{\ms{S}}_{\dot{2}} \;,
\end{align}
where
\begin{align}
  \{\ms{Q}, \ms{S}\} = 2\mr{i}r^{-1} (\ms{J}_{3\vphantom{\dot{1}}} - \ms{R}_{\dot{1}\dot{2}}) \;.
\end{align}
Indeed, the BPS operators we will consider are closed with respect to both $\ms{Q}$ and $\ms{S}$. Furthermore, the supercharge we employ for localization is
\begin{align}\label{EQ:localizing_supercharge}
  \mathbf{Q} = \f{\mr{i}}{2}(\ms{Q} - \ms{S}) \;,
\end{align}
squaring to (a multiple of) the twisted rotation $\ms{J}_3 - \ms{R}_{\dot{1}\dot{2}}$. The particular linear combination adopted in \eqref{EQ:localizing_supercharge} has been chosen for later convenience, however, since $\ms{Q}$ and $\ms{S}$ are both nilpotent, any other linear combination between them would be equivalent. Since Higgs-branch operators are singlets under $\mf{su}(2)_{\mr{C}}$, their BPS locus is then located at the fixed points of the isometry $\ms{J}_3$, namely the north and south pole of $S^2$.

As the next step, we need to bring the localizing supercharge off shell. As mentioned in Section \ref{SEC:setup} this can be done by identifying an appropriate set of auxiliary spinors $\nu^{a\dot{a}}$ that satisfies the constraints \eqref{EQ:auxiliary_identities}. For any supercharge in \eqref{EQ:su(1|1)_generators}, these constraints are solved by
\begin{align}\label{EQ:auxiliary_solution}
  \nu^{a\dot{a}} = -(\sigma^3)^{\dot{a}}\vphantom{)}_{\dot{c}}\,\xi^{a\dot{c}} \;.
\end{align}

The choice of $\textbf{Q}$ as localizing supercharge fixes the form of the Killing vector
\begin{align}
  v &= - r^{-1}\,\partial_{\varphi} \;,
\end{align}
which generates the isometry that, in turn, determines the localizing submanifold $\mh{M}_v$, as in Section \ref{SEC:localization}. We find that $v$ vanishes at the north and south poles of $S^2$, and as such, $\mh{M}_v$ is given by two copies of $S^1$, that we will denote with $S^1_{\mr{N}}$ and $S^1_{\mr{S}}$. These are located, respectively, at $\theta=0$ and $\theta=\pi$.
The choice of localizing supercharge also determines the form of the scalars
\begin{align}
  X_{ac} &= \f{1}{2}\begin{pmatrix} 1&\cos\theta \\ \cos\theta&1 \end{pmatrix} \;, &
  \bar{X}_{\dot{a}\dot{c}} &= \f{\sin\theta}{2}\begin{pmatrix} -\mr{i}e^{-\mr{i}\varphi}&0 \\ 0&\mr{i}e^{\mr{i}\varphi} \end{pmatrix} \;,
\end{align}
and the one-forms
\begin{align}
  w &= \frac{\beta}{2}\cos\theta\,\mr{d}t \;, \cr
  \Lambda_{ac} &= -\f{\sigma}{2}\begin{pmatrix}
    \beta\cos\theta\,\mr{d}t+\sin\theta\,\mr{d}\theta&\beta\mr{d}t \\
    \beta\mr{d}t&\beta\cos\theta\,\mr{d}t-\sin\theta\,\mr{d}\theta
  \end{pmatrix} \;.
\end{align}
that enter in the definition of the one-form Lagrangian $\alpha_1$, as in \eqref{EQ:one_form_action}.

\begin{figure}[ht]
\centering
\begin{tikzpicture}[thick,scale=1.3,>=latex]
\begin{scope}[xshift=-2.1cm]
  \path[use as bounding box] (-1.6,-1.9) rectangle (1.6,1.9);
  \colorlet{shadow}{black!10}
  \begin{scope}
    \clip (-1.5,-1.5) rectangle (1.5,1.5);
    \fill [shadow] (0,0) circle[radius=1.5];
    \fill [white] (-0.1,0.5) circle[radius=1.7];
  \end{scope}
  \draw[->-=0.46, black!50] (0,1.25) [partial ellipse=-270:+90:0.57 and 0.25];
  \draw[->-=0.4, black!50] (0,0.97) [partial ellipse=-200:20:1.05 and 0.48];
  \draw[->-=0.39, black!50] (0,0.51) [partial ellipse=-190:10:1.39 and 0.6];
  \draw[->-=0.38, black!50] (0,0) [partial ellipse=-180:0:1.5 and 0.68];
  \draw[->-=0.38, black!50] (0,-0.52) [partial ellipse=-170:-10:1.39 and 0.63];
  \draw[ultra thick, black] (0,0) circle (1.5);
  \fill [black](0,1.35) ellipse (1.5pt and 1pt);
  \fill [black!30](0,-1.35) ellipse (1.5pt and 1pt);
\end{scope}
\end{tikzpicture}
\qquad
\begin{tikzpicture}[thick,scale=1.3]
  \node at (-2.5,0) {$\longmapsto$};
  \colorlet{shadow}{black!5}
  \begin{scope}
    \clip (-1.5,-1.5) rectangle (1.5,1.5);
    \fill [shadow] (0,0) circle[radius=1.5];
    \fill [white] (-0.1,0.5) circle[radius=1.7];
  \end{scope}
  \draw[->-=0.46, black!10] (0,1.25) [partial ellipse=-270:+90:0.57 and 0.25];
  \draw[->-=0.4, black!10] (0,0.97) [partial ellipse=-200:20:1.05 and 0.48];
  \draw[->-=0.39, black!10] (0,0.51) [partial ellipse=-190:10:1.39 and 0.6];
  \draw[->-=0.38, black!10] (0,0) [partial ellipse=-180:0:1.5 and 0.68];
  \draw[->-=0.38, black!10] (0,-0.52) [partial ellipse=-170:-10:1.39 and 0.63];
  \draw[ultra thick, black!20] (0,0) circle (1.5);
  \draw[ultra thick, dashed, red] (0,1.35) -- (0,-1.35);
  \fill [black](0,1.35) ellipse (1.5pt and 1pt);
  \fill [black](0,-1.35) ellipse (1.5pt and 1pt);
  \draw[ultra thick, black] (3.0, 1.0) ellipse (1.0 and 0.5);
  \draw[ultra thick, black] (3.0,-1.0) ellipse (1.0 and 0.5);
  \node at (3.6, 0.95) {$S^1_{\mathrm{N}}$};
  \node at (3.6,-0.95) {$S^1_{\mathrm{S}}$};
\draw[green!50!black!50,->] (0.3, 1.6) to[out=25, in=145] (2.0, 1.4);
\draw[green!50!black!50,->] (0.3,-1.6) to[out=-25, in=-145] (2.0,-1.4);
\end{tikzpicture}
\caption{A graphical representation of the quotient $\mh{M}_v = (S^2 \times S^1)/\mr{U}(1)$, where the $\mr{U}(1)$ acts as the azimuthal rotation generated by $\partial_\varphi$. In the picture, every point on the sphere has an asssociated $S^1$ fiber. The fixed point locus $\mh{M}_v$ is the disjoint union of the two $S^1$ fibers located at the poles of the two-sphere.}
\end{figure}
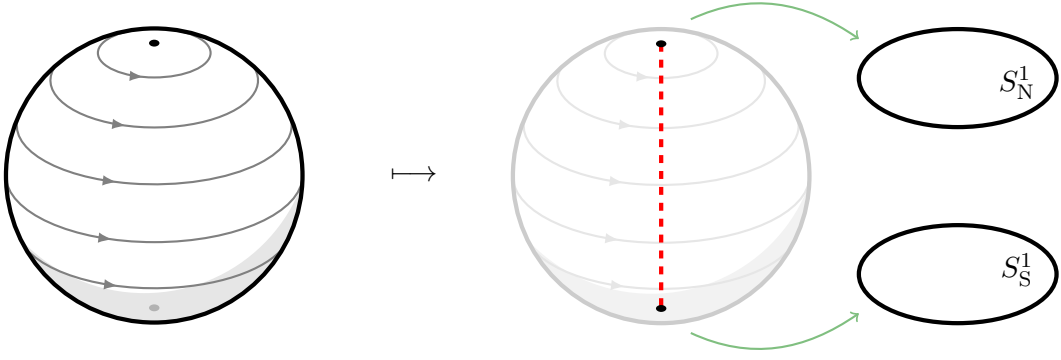

On the two copies of $S^1$, at the north and the south poles of $S^2$, we find
\begin{align}\label{EQ:1d_forms_S2xS1}
  i_{\mr{N}}^*\,\alpha_1 &= 1/2\,(\tilde{q}^+\mh{D}q^+ - \mh{D}\tilde{q}^+q^+) \;, \cr
  i_{\mr{S}}^*\,\alpha_1 &= 1/2\,(\tilde{q}^-\overline{\mh{D}}q^- - \overline{\mh{D}}\tilde{q}^-q^-) \;,
\end{align}
where
\begin{align}
  q^{\pm} &= q^1\pm q^2 \;, & \tilde{q}^{\pm} &= \tilde{q}^1\pm\tilde{q}^2 \;,
\end{align}
and
\begin{align}
  \mh{D} &= (\partial_t - \mr{i}\zeta^{\phantom{*}})\,\mr{d}t \,, \cr
  \overline{\mh{D}} &= (\partial_t - \mr{i}\zeta^*)\,\mr{d}t \,.
\end{align}
Since
\begin{align}
  \mr{e}_v(NS^1_{\mr{N}}) = - \mr{e}_v(NS^1_{\mr{S}}) = \f{1}{2\pi r} \;,
\end{align}
we can simply integrate by parts the two expressions in \eqref{EQ:1d_forms_S2xS1} and find
\begin{align}\label{EQ:1d_action_S2xS1}
  S_{1\mr{d}} = 2\pi r\oint_{S^1_{\mr{N}}}\tilde{q}^+\mh{D}q^+ - 2\pi r\oint_{S^1_{\mr{S}}}\tilde{q}^-\overline{\mh{D}}q^-
\end{align}

On $\mh{M}_v = S^1_{\mr{N}}\cup S^1_{\mr{S}}$, where $v$ vanishes, one can find local BPS operators, since the action of $\mc{L}_v$ is trivial. One can set to zero the variation of a generic linear combination of scalars,
\begin{align}
  0 &= \delta(b_aq^{a}) = b_a\xi^{a\dot{a}}\psi_{\dot{a}} \;, \cr
  0 &= \delta(\tilde{b}_a\tilde{q}^{a}) = \tilde{b}_a\xi^{a\dot{a}}\tilde{\psi}_{\dot{a}} \;,
\end{align}
and look for nontrivial solutions for the coefficients $b_a$, $\tilde{b}_a$.
If such solutions exist, these must be null eigenvectors of $X^{ac}$, which in fact  has vanishing determinant on $\mh{M}_v$, as noted in \eqref{EQ:determinant_formula}. In the present case, a quick computation shows that the BPS operators on $S^1_{\mr{N}}$ and $S^1_{\mr{S}}$ are precisely the $q^+$, $\tilde{q}^+$ and $q^-$, $\tilde{q}^-$, respectively, that appear in the one-dimensional action \eqref{EQ:1d_action_S2xS1}. The R-symmetry structure of these operators is determined by the choice of $\ms{Q}$ and $\ms{S}$ in \eqref{EQ:su(1|1)_generators}. A different choice would lead to a different set of BPS operators.

We conclude this section with a comment on the localization argument. Given our discussion of Section \ref{SEC:localization}, one might be tempted to consider a localizing supercharge that would close on a different isometry, namely the isometry $\ms{Z}$ that generates translations along the $S^1$ coordinate $t$. However, one is faced immediately with an apparent contradiction. In fact, such an isometry has vanishing fixed point locus. This, because of \eqref{EQ:localization_formula}, seems to imply that the action vanishes on the associated BPS solutions, thus leaving us with a trivial localizing theory. The problem with this choice if isometry is that it leads to a one-form $w$ that is not well defined on the entire manifold $\mh{M}$, as one can explicitly check by direct computation. With $\alpha_1$ not globally defined of $\mh{M}$, the localization theorem that leads to \eqref{EQ:localization_formula} does not hold.

\section{One-Dimensional Theory}\label{SEC:1d_theory}
The one-dimensional action $S_{\mr{1d}}$ defined in \eqref{EQ:1d_action_S2xS1} determines a quantum mechanics with path integral
\begin{align}\label{EQ:1d_path_integral}
  Z_{1\mr{d}} = \int_{\mr{BPS}}[\mr{d}\tilde{q}^+_{\mr{N}}][\mr{d}q^+_{\mr{N}}][\mr{d}\tilde{q}^-_{\mr{S}}][\mr{d}q^-_{\mr{S}}] \; e^{-S_{1\mr{d}}} \;.
\end{align}
This one-dimensional theory captures the correlators of Higgs branch operators built out of $\tilde{q}^+_{\mr{N}}$, $q^+_{\mr{N}}$, $\tilde{q}^-_{\mr{S}}$, $q^-_{\mr{S}}$. By this we mean that such correlators are directly given by the Green's function descending from \eqref{EQ:1d_action_S2xS1}, which in Fourier space read
\begin{align}
  \tilde{\mc{G}}_{+,k} \sim \f{1}{k+\zeta}
\end{align}
for $\tilde{q}^+_{\mr{N}},q^+_{\mr{N}}$ and
\begin{align}
  \tilde{\mc{G}}_{-,k} \sim \f{1}{k+\zeta^*}
\end{align} for $\tilde{q}^-_{\mr{S}}$, $q^-_{\mr{S}}$. In this section we shall check this statement explicitly via different methods.

The functional weight of the quantum mechanics path integral in \eqref{EQ:1d_action_S2xS1} is  the exponential of a simple Gaussian action, but it requires some care. For instance, it is not obviously positive definite because only first-order derivatives appear in it. The BPS solutions compatible with the reality conditions \eqref{EQ:vector_reality_conditions} select a middle-dimensional integration cycle whose parametrization can be found in Appendix \ref{APP:BPS_hyper}, where the BPS configurations are obtained explicitly. These are given in terms of two independent set of complex Fourier coefficients.

For simplicity of notation, we will mainly work with the case of a single hypermultiplet ($N=1$) and $G\simeq\mr{U}(1)$. The extensions to more general cases is straightforward and will be discussed later in this section and in Section \ref{SEC:correlators}. Moreover, from now on we will drop the subscript $\mr{N}$/$\mr{S}$, as the superscript $\pm$ is sufficient to resolve the ambiguity.

By plugging the solutions \eqref{EQ:BPS_sols_hyper} in the one-dimensional action \eqref{EQ:1d_action_S2xS1}, one finds\footnote{We have dropped the subscript $+$, so the coefficients $u_k$ and $v_k$ that appear in the expression for $S_{1\mr{d}}$ are simply $u_{+,k}$ and $v_{+,k}$ of Appendix \ref{APP:BPS_hyper}.}
\begin{align}
  S_{1\mr{d}} = 4\pi^2r\sum_{k\in\mb{Z}}|k+\zeta|\,\Big[\big(|u^{\vphantom{*}}_{k}|^2+|v^{\vphantom{*}}_{k}|^2\big)\sinh(2\omega_k)-\mr{i}\big(u^{\vphantom{*}}_{k}v^*_{k}+u^*_{k}v^{\vphantom{*}}_{k}\big)\Big] \;,
\end{align}
whose real part is manifestly positive semidefinite.
The partition function is obtained as a product of complex Gaussian integrals over each Fourier mode,
\begin{align}\label{EQ:1d_partiton_function}
  Z_{1\mr{d}}(\zeta,\zeta^*) &= \prod_{k\in\mb{Z}}\int_{\mb{C}^2}\mr{d}u_k^{\vphantom{*}}\,\mr{d}u_k^*\,\mr{d}v_k^{\vphantom{*}}\,\mr{d}v_k^* \; \cosh^{2}(2\omega_k) \; e^{-\textbf{z}_k^\dagger\textbf{M}^{\vphantom{\dagger}}_k\textbf{z}^{\vphantom{\dagger}}_k} \cr
  &= \prod_{k\in\mb{Z}} \f{1}{4\pi^2r^2|k+\zeta|^2} \;,
\end{align}
where
\begin{align}
  \textbf{z}_k &= \begin{pmatrix}u_k\\v_k\end{pmatrix} \;, & \textbf{M}_k &= 4\pi^2r|k+\zeta|\begin{pmatrix}\sinh(2\omega_k)&-\mr{i}\\-\mr{i}&\sinh(2\omega_k)\end{pmatrix} \;.
\end{align}
The Jacobian $\cosh^2(2\omega_k)$ cancels an analogous term coming from the determinant of $\mathbf{M}_k$.

From the inverse of $\textbf{M}_k$ we read off the correlators
\begin{align}
  \langle u^*_{k}\,u^{\vphantom{*}}_{k'} \rangle &= \langle v^*_{k}\,v^{\vphantom{*}}_{k'} \rangle = \delta_{k,k'}^{\vphantom{*}} \f{\sinh(2\omega_k)}{4\pi^2r|k+\zeta|\cosh^2(2\omega_k)} \;, \cr
  \langle u^*_{k}\,v^{\vphantom{*}}_{k'} \rangle &= \langle v^*_{k}\,u^{\vphantom{*}}_{k'} \rangle = \delta_{k,k'}^{\vphantom{*}} \f{\mr{i}}{4\pi^2r|k+\zeta|\cosh^2(2\omega_k)} \;.
\end{align}
According to the solutions \eqref{EQ:BPS_sols_hyper}, the above combine into
\begin{align}\label{EQ:1d_correlators}
  \langle \tilde{q}^+(t_2) \, q^+(t_1) \rangle &= +\f{\mr{i}}{4\pi^2r}\sum_{k\in\mb{Z}} \frac{1}{k+\zeta^{\phantom{\star}}} \, e^{\mr{i}k(t_2-t_1)} \;, \cr
  \langle \tilde{q}^-(t_2) \, q^-(t_1) \rangle &= -\f{\mr{i}}{4\pi^2r}\sum_{k\in\mb{Z}} \frac{1}{k+\zeta^*} \, e^{\mr{i}k(t_2-t_1)} \;.
\end{align}
These correlators coincide with the Green functions obtained from \eqref{EQ:1d_action_S2xS1} by naive Fourier expansion in $t$. This fact is no coincidence and will be explored in Section \ref{SEC:Morse_cycle}.
When performing the sum, one finds that
\begin{align}
  \langle \tilde{q}^{\pm}(t_2) \, q^{\pm}(t_1) \rangle = \mc{G}_{\pm}(t_2-t_1) \;,
\end{align}
where
\begin{align}\label{EQ:1d_correlators_G}
  \mc{G}_{+}(t) &= -\f{\sign(t)-\mr{i}\cot(\pi\zeta)}{4\pi r}\,e^{-\mr{i}\zeta t} \;, \cr
  \mc{G}_{-}(t) &= +\f{\sign(t)-\mr{i}\cot(\pi\zeta^*)}{4\pi r}\,e^{-\mr{i}\zeta^*t} \;.
\end{align}
To be more precise, the expressions above are only valid in the range allowed for the coordinates $t_1, t_2 \in [0,2\pi)$, i.e.\ for $-2\pi<t<2\pi$, with $t=t_2-t_1$. For values of $t$ outside of this domain, $\mc{G}_+$ and $\mc{G}_-$ should be defined by simply enforcing the periodicity condition
\begin{align}
  \mc{G}_{\pm}(t) = \mc{G}_{\pm}(t+2\pi) \;,
\end{align}
which is manifest in \eqref{EQ:1d_correlators}.

Let us now see how this generalizes to $N>1$. The partition function in \eqref{EQ:1d_partiton_function} becomes
\begin{align}
  Z_{1\mr{d}}(\zeta,\zeta^*) &= \prod_{k\in\mb{Z}} \prod_{\rho\in\mc{R}} \f{1}{4\pi^2|k+\rho(\zeta)|^2} \;,
\end{align}
where we have set $r=1$ and the product over $\rho$ is taken over the weights of the representation $\mc{R}$.
We can give meaning to the product over the Fourier modes by means of zeta-function regularization. Specifically, since for any $a>0$ and $b\in\mb{C}$ this prescribes
\begin{align}\label{EQ:zeta_regularized_product}
  \prod_{k\in\mb{Z}} |ak+b|^2 = 4|\sin(\pi b/a)|^2 \;,
\end{align}
then,
\begin{align}\label{EQ:1d_partiton_function_full}
  Z_{1\mr{d}}(\zeta,\zeta^*) &= \prod_{\rho\in\mc{R}} \f{1}{4|\sin(\pi \rho(\zeta))|^2} \;.
\end{align}
This is indeed the result that we find in Appendix \ref{APP:3d_partition_function}, where we perform a direct computation of the partition function of a three-dimensional hypermultiplet.

When $\sigma=0$, we find that \eqref{EQ:1d_partiton_function_full} coincides with $\mc{I}^{\mr{C}}$, which is the so-called C-twist limit of the three-dimensional $\mc{N}=4$ superconformal index $\mc{I}$ computed e.g.\ in \cite{Razamat:2014pta}. A limit of a supersymmetric index specifically counting Higgs branch local operators was explored also in \cite{Gaiotto:2019jvo,Okazaki:2019ony}.
Explicitly, for a massless hypermultiplet coupled to a $U(1)$ background connection $a$, we have 
\begin{equation}
  \mc{I}^{\mr{C}}(\tilde{x}, z) = \tr_{\mc{H}[S^2]}(-1)^F \tilde{x}^{E - R_{\mc{H}}} z^{J_f} \Big|_{\mathfrak m=0} = \f{1}{(1-z)(1-z^{-1})} \;.
\end{equation}
where $z=e^{2\pi\mr{i}a}$ is the flavor fugacity for $a$, $J_f$ is the generator of the $U(1)$ flavor symmetry and the trace is over the Hilbert space $\mathcal H[S^2]$ of states on $S^2$. By setting $\mf{m}=0$ we are neglecting monopole configurations that do not appear in our analysis as discussed in Section \ref{SEC:S2S1_background}. Here, we observe that the Higgs branch local operators with topological correlation functions represent the only contribution to $\mc{I}^{\mr{C}}(\tilde{x}, z)$. This is analogous to what happens in four dimensions, where the Higgs branch local operators with holomorphic correlators are the main contribution to the Schur limit of the superconformal index \cite{Beem:2013sza,Pan:2019bor}.

\subsection{Morse Theory Approach}\label{SEC:Morse_cycle}
So far, we have shown how the one-dimensional action \eqref{EQ:1d_action_S2xS1} gives rise to a Gaussian quantum mechanics that effectively reproduces the partition function and the correlators of a hypermultiplet on $S^2\times S^1$.
However, the path integral \eqref{EQ:1d_path_integral} is well defined only once a middle-dimensional integration cycle is specified. A generic cycle would not even lead to a convergent integration. The correct cycle is not obtained by simply specifying the reality conditions in \eqref{EQ:hyper_reality_conditions}. In fact, from the point of view of the three-dimensional theory, the fields over which the functional integration in \eqref{EQ:1d_path_integral} is performed are four independent complex degrees of freedom. The additional constraints imposed in \eqref{EQ:1d_partiton_function} come from the explicit form of the solutions of the BPS equations \eqref{EQ:hyper_bps} and determine a nontrivial integration cycle for the path integral.
Consequently, one would be led to conclude that, although the localization formula \eqref{EQ:localization_formula} is valid on any generic background, one would still have to solve the BPS equations for the case at hand in order to give meaning to the resulting one-dimensional theory. However, as already noticed in \cite{Dedushenko:2016jxl}, it is possible to make sense of the functional integral in \eqref{EQ:1d_action_S2xS1} by directly addressing the problem of its convergence. We adopt the approach described in \cite{Witten:2010zr} and we refer the reader to that reference for a full exposition of the procedure.

We start by considering the space of complex fields in the path integral \eqref{EQ:1d_path_integral}, i.e.\ the space of maps $S^1\to\mb{C}^4$. First, we split the fields in terms of their real and imaginary parts with
\begin{align}\label{EQ:Morse_fields}
  q^\pm &= x^\pm + \mr{i}y^\pm \;, &
  \tilde{q}^\pm &= \tilde{x}^\pm + \mr{i}\tilde{y}^\pm \;.
\end{align}
Then, we define a Morse function $h$ as the real part of the one-dimensional action \eqref{EQ:1d_action_S2xS1} (up to an overall negative real constant),
\begin{align}
  h = h_{\mr{N}} + h_{\mr{S}} \;,
\end{align}
where
\begin{align}\label{EQ:Morse_function}
  h_{\mr{N}} &= -\oint \mr{d}t \; [\tilde{x}^+\partial_tx^+ - \tilde{y}^+\partial_ty^+ + a (\tilde{x}^+y^+ + \tilde{y}^+x^+) + \sigma (\tilde{y}^+y^+ - \tilde{x}^+x^+)] \;, \cr
  h_{\mr{S}} &= -\oint \mr{d}t \; [\tilde{y}^-\partial_ty^- - \tilde{x}^-\partial_tx^- - a (\tilde{x}^-y^- + \tilde{y}^-x^-) + \sigma (\tilde{y}^-y^- - \tilde{x}^-x^-)] \;.
\end{align}

We start by identifying the critical points of $h$, i.e. the solutions of $\delta h=0$.
For a generic value of $\zeta$, $h$ has a single critical point where all fields are identically vanishing.
The idea is to define the integration cycle by considering the gradient flow lines originating from the critical point. If the Morse index\footnote{The Morse index is defined as the number of negative eigenvalues of the Hessian matrix.} of $h$ at the critical point is half the dimension of the target space, the cycle is indeed middle-dimensional. The convergence of the functional integration is guaranteed by the fact that $h$ decreases along the flow.

If we collectively denote the fields in \eqref{EQ:Morse_fields} with $Y^A$, we can write the flow equations as
\begin{align}
  \f{\partial Y^A}{\partial s} = -g^{AB}\f{\delta h}{\delta Y^B} \;,
\end{align}
where $s\in(-\infty,0]$ is the flow parameter.
The gradient flow depends on the choice of a target-space metric $g_{AB}$. By choosing the canonical metric on $\mb{C}^4$
we obtain the set of equations
\begin{align}
  \partial_s x^+ &= -\partial_t \tilde{x}^+ + a \tilde{y}^+ - \sigma \tilde{x}^+ \;, &
  \partial_s x^- &= +\partial_t \tilde{x}^- - a \tilde{y}^- - \sigma \tilde{x}^- \;, \cr
  \partial_s y^+ &= +\partial_t \tilde{y}^+ + a \tilde{x}^+ + \sigma \tilde{y}^+ \;, & 
  \partial_s y^- &= -\partial_t \tilde{y}^- - a \tilde{x}^- + \sigma \tilde{y}^- \;, \cr
  \partial_s \tilde{x}^+ &= +\partial_t x^+ + a y^+ - \sigma x^+ \;, &
  \partial_s \tilde{x}^- &= -\partial_t x^- - a y^- - \sigma x^- \;, \cr
  \partial_s \tilde{y}^+ &= -\partial_t y^+ + a x^+ + \sigma y^+ \;, &
  \partial_s \tilde{y}^- &= +\partial_t y^- - a x^- + \sigma y^- \;,
\end{align}
which, when recasted in terms of the original fields and their complex conjugates, read
\begin{align}
  -\partial_s q^+ &= (\partial_t - \mr{i}\zeta^*) (+\tilde{q}^+)^* \;, &
  -\partial_s q^- &= (\partial_t - \mr{i}\zeta) (-\tilde{q}^-)^* \;, \cr
  -\partial_s \tilde{q}^+ &= (\partial_t + \mr{i}\zeta^*) (-q^+)^* \;, &
  -\partial_s \tilde{q}^- &= (\partial_t + \mr{i}\zeta) (+q^-)^* \;.
\end{align}

As mentioned, we look for solutions that originate from the critical point, i.e. solutions   vanishing for $s\rightarrow-\infty$. These are
\begin{align}
  q^+(s;t) &= \sum_{k\in\mb{Z}} b_k \, e^{-\mr{i}kt+\omega_ks} \;, &
  \tilde{q}^+(s;t) &= +\mr{i}\sum_{k\in\mb{Z}} \f{k+\zeta^*}{|k+\zeta|} \, b^*_k \, e^{+\mr{i}kt+\omega_ks} \;, \cr
  q^-(s;t) &= \sum_{k\in\mb{Z}} d_k \, e^{-\mr{i}kt+\omega_ks} \;, &
  \tilde{q}^-(s;t) &= -\mr{i}\sum_{k\in\mb{Z}} \f{k+\zeta}{|k+\zeta|} \, d^*_k \, e^{+\mr{i}kt+\omega_ks} \;.
\end{align}
It is convenient to pick the solutions ``at the end of the flow'', where $s=0$, although any other choice will do, as it amounts to a simple rescaling of the coefficients $b_k$ and $d_k$.

In terms of these solutions, the one-dimensional action reads
\begin{align}
  S_{\mr{Morse}} = 4\pi^2r\sum_{k\in\mb{Z}}(|b_k|^2+|d_k|^2)|k+\zeta| \;,
\end{align}
and the partition function, in analogy with what we have done in the previous section, can be computed with
\begin{align}\label{EQ:1d_partiton_function_Morse}
  Z_{1\mr{d}}(\zeta,\zeta^*) &= \prod_{k\in\mb{Z}}\int_{\mb{C}^2}\mr{d}b_k^{\vphantom{*}}\,\mr{d}b_k^*\,\mr{d}d_k^{\vphantom{*}}\,\mr{d}d_k^* \; e^{-\textbf{z}_k^\dagger\textbf{M}^{\vphantom{\dagger}}_k\textbf{z}^{\vphantom{\dagger}}_k} \;,
\end{align}
where
\begin{align}
  \textbf{z}_k &= \begin{pmatrix}b_k\\d_k\end{pmatrix} \;, & \textbf{N}_k &= 4\pi^2r|k+\zeta|\begin{pmatrix}1&\;0\\0&\;1\end{pmatrix} \;.
\end{align}
This integration cycle has the additional benefit of diagonalizing the action, thus making manifest the factorization of the theory into two noninteracting ``north'' and ``south'' contributions.
Notice how \eqref{EQ:1d_partiton_function_Morse} gives the same result of \eqref{EQ:1d_partiton_function}. Likewise, one finds that the correlators
\begin{align}
  \langle b^*_k\,b^{\vphantom{*}}_{k'} \rangle = \langle d^*_k\,d^{\vphantom{*}}_{k'} \rangle &= \f{\delta_{k,k'}}{4\pi^2r|k+\zeta|} \;, \cr
  \langle b^*_k\,d^{\vphantom{*}}_{k'} \rangle = \langle d^*_k\,b^{\vphantom{*}}_{k'} \rangle &= 0 \;,
\end{align}
give rise to correlators $\langle \tilde{q}^\pm(t_2)\,q^\pm(t_1) \rangle$ that are identical to the ones computed in \eqref{EQ:1d_correlators}.

\section{Topological Correlators and Flavor Symmetry Enhancement}\label{SEC:correlators}
\subsection{Noncommutative Star Product}
We will now look at correlation functions of composite BPS operators. We will consider the case where $N$ is generic. In doing so, we explicit the flavor indices carried by the fields appearing in the one-dimensional theory, i.e.\ we write $\tilde{q}^+{}_i$, $q^{+\,j}$, $\tilde{q}^-{}_k$, $q^{-\,j}$, where raised and lowered indices $i,j,k,l=1,\ldots, N$ are, respectively, fundamental and anti-fundamental $\mr{U}(N)$ indices.

The presence of a background $G$-connection, also referred to as ``weak gauging'', breaks the flavor subgroup $\mr{U}(N)$ down to the normalizer of $G$ in $\mr{U}(N)$. In this section and in the following, we will take $G\simeq\mr{U}(1)$, thus obtaining a residual $\mr{SU}(N)$ flavor group.
In terms of flavor indices, the correlators
\begin{align}
  \langle \tilde{q}^{\pm}{}_j(t_2) \, q^{\pm\,i}(t_1) \rangle = \mc{G}_{\pm}{}^i{}_j(t_2-t_1) \;,
\end{align}
are diagonal and can be obtained directly from the correlators in \eqref{EQ:1d_correlators_G} computed for $N=1$. We write them as
\begin{align}\label{EQ:1d_correlators_with_flavor}
  \mc{G}_{+}{}^i{}_j(t) &= \bigg({-}\f{\delta^i{}_j\sign(t)}{4\pi r}+\hat{\mc{G}}_{+}{}^i{}_j\bigg)\,e^{-\mr{i}\zeta t} \;, \cr
  \mc{G}_{-}{}^i{}_j(t) &= \bigg({+}\f{\delta^i{}_j\sign(t)}{4\pi r}+\hat{\mc{G}}_{-}{}^i{}_j\bigg)\,e^{-\mr{i}\zeta^*t} \;,
\end{align}
where we have defined
\begin{align}\label{EQ:topological_same_t_correlators}
  \hat{\mc{G}}_{+}{}^i{}_j &= +\delta^i{}_j\,\f{\mr{i}\cot(\pi\zeta)}{4\pi r} \;, \cr
  \hat{\mc{G}}_{-}{}^i{}_j &= -\delta^i{}_j\,\f{\mr{i}\cot(\pi\zeta^*)}{4\pi r} \;.
\end{align}
The correlators $\langle \tilde{q}^{\pm}{}_j(t_2) \, q^{\pm\,i}(t_1) \rangle$ are discontinuous at coincident insertion points. When $t_1=t_2$ one finds\footnote{
  This comes from regularizing the Fourier sum in \eqref{EQ:1d_correlators} with
  \begin{align*}
    \lim_{M\to\infty} \sum_{k=-M}^M \f{1}{k+\zeta} = \pi \cot(\pi\zeta) \;.
  \end{align*}
}
\begin{align}
  \mc{G}_{+}{}^i{}_j(0) &= \hat{\mc{G}}_{+}{}^i{}_j\,e^{-\mr{i}\zeta t} \;, &
  \mc{G}_{-}{}^i{}_j(0) &= \hat{\mc{G}}_{-}{}^i{}_j\,e^{-\mr{i}\zeta^*t} \;,
\end{align}
which amount to simply enforcing $\sign(0) = 0$ in \eqref{EQ:1d_correlators_with_flavor}.

We now consider correlators of local operators $\mc{O}(t)$ defined as polynomials in $\tilde{q}^{\pm}{}_i$ and $q^{\pm\,j}$. Without loss of generality we will consider only $t$-ordered correlators, i.e. correlators
\begin{align}
  \langle \mc{O}_n(t_n) \, \ldots \, \mc{O}_2(t_2) \, \mc{O}_1(t_1) \rangle
\end{align}
where $t_1 < t_2 < \ldots t_n$. An interesting property of these correlators is that their dependence on the coordinates $t_m$ is trivial. In fact, when performing Wick contraction one finds that all terms share a common exponential prefactor that encapsulates the overall $t$-dependence of the correlator. A similar factorization was also considered in \cite{Dedushenko:2016jxl}. We can extract this term with
\begin{align}
  \langle \mc{O}_n(t_n) \, \ldots \, \mc{O}_2(t_2) \, \mc{O}_1(t_1) \rangle = e^{\mr{i}\sum_m\!\mc{R}_m(\zeta)\,t_m} \; \langle\!\langle \mc{O}_n \, \ldots \, \mc{O}_2 \, \mc{O}_1 \rangle\!\rangle \;,
\end{align}
where $\mc{O}_m$ belongs to the $\mc{R}_m$ representation of $G$ and $\langle\!\langle \ldots \rangle\!\rangle$ is a topological correlator. The latter is defined from the original correlator by setting $t_{m+1} = t_{m}+\epsilon$ and then taking the limit for $\epsilon\to 0^+$.

\begin{figure}[ht]
  \centering
  \begin{tikzpicture}[thick,scale=0.8]
    \begin{scope}[xshift=-5cm]
      \draw[ultra thick, black] (0.0, 0.0) ellipse (3.0 and 1.5);
      \node at (3.1,1.4) {$S^1_{\mathrm{N/S}}$};
      \node[black!50,rotate=-40] at (-3.,-1.1) {\textbf{\ldots}};
      \node[black!50,rotate=+40] at (+3.,-1.1) {\textbf{\ldots}};
      \fill [black](-1.8,-1.2) circle (4pt) node[below, xshift=-5, yshift=-3] {$\mc{O}_{m-1}$};
      \fill [black](-0.62,-1.47) circle (4pt) node[below, xshift=0, yshift=-5] {$\mc{O}_{m}$};
      \fill [black](+0.62,-1.47) circle (4pt) node[below, xshift=5, yshift=-5] {$\mc{O}_{m+1}$};
      \fill [black](+1.8,-1.2) circle (4pt) node[below, xshift=15, yshift=-3] {$\mc{O}_{m+2}$};
      \draw[thick, -latex] (0,0) [partial ellipse=60:120:2.6 and 1.2];
      \node at (0,0.9) {$t$};
    \end{scope}
      \node at (0,0) {$\longmapsto$};
    \begin{scope}[xshift=+5cm]
      \draw[ultra thick, black] (0.0, 0.0) ellipse (3.0 and 1.5);
      \node at (-3.1,1.4) {$S^1_{\mathrm{N/S}}$};
      \node[black!50,rotate=-40] at (-3.,-1.1) {\textbf{\ldots}};
      \node[black!50,rotate=+40] at (+3.,-1.1) {\textbf{\ldots}};
      \fill [black](-1.8,-1.2) circle (4pt) node[below, xshift=-5, yshift=-3] {$\mc{O}_{m-1}$};
      \fill [black](0,-1.5) circle (4pt) node[below, xshift=3, yshift=-4] {$(\mc{O}_{m+1}*\mc{O}_{m})$};
      \fill [black](+1.8,-1.2) circle (4pt) node[below, xshift=15, yshift=-3] {$\mc{O}_{m+2}$};
      \draw[thick, -latex] (0,0) [partial ellipse=60:120:2.6 and 1.2];
      \node at (0,0.9) {$t$};
    \end{scope}
    \end{tikzpicture}
  \caption{In a topological correlator, two adjacent insertions can be replaced by a single one corresponding to their star product.}\label{FIG:star_product}
\end{figure}
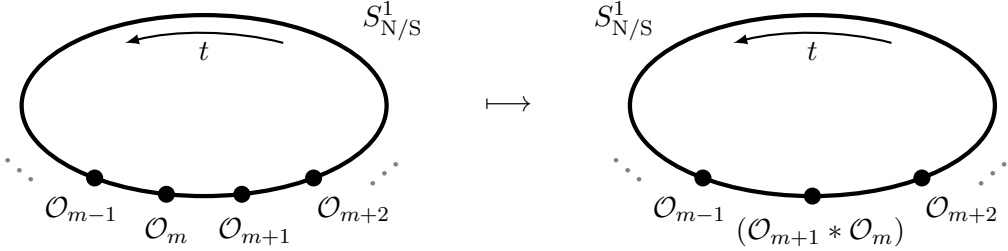

We can therefore think of the quantum mechanics in \eqref{EQ:1d_path_integral} as defining two noninteracting disconnected topological theories, on $S^1_{\mr{N}}$ and $S^1_{\mr{S}}$.
These topological theories have an associated operator algebra which is naturally defined through a noncommutative star product $\ast$. See also \cite{Beem:2016cbd} for an analogous construction in the case of superconformal field theories. We will now explain how this comes about. Let us consider a topological correlator $\langle\!\langle \mc{O}_n \, \ldots \, \mc{O}_2 \, \mc{O}_1 \rangle\!\rangle$. One can obtain an equivalent correlator by using the star product to fuse any two distinct adjacent operators $\mc{O}_{m+1}$ and $\mc{O}_m$ into a single local insertion
\begin{align}
  \mc{O}'_{m} = \mc{O}^{\vphantom{'}}_{m+1}\ast\mc{O}^{\vphantom{'}}_{m} \;,
\end{align}
which preserves its topological order with respect to all other insertions $\mc{O}_{m'}$. This is illustrated in Figure \ref{FIG:star_product}.

One can recursively apply the star product to reduce any topological correlator to the one-point function of a certain polynomial in $\tilde{q}^\pm{}_i$ and $q^{\pm\,j}$. This one-point function can be computed by performing Wick contractions with the correlators $\hat{\mc{G}}_{\pm}{}^i{}_j$ as defined in \eqref{EQ:topological_same_t_correlators}.
The product between any two operators can be generated starting from the fundamental identities
\begin{align}\label{EQ:star_product_qq}
  \tilde{q}^{\pm}{}_j \ast q^{\pm\,i} &= \tilde{q}^{\pm}{}_j \, q^{\pm\,i} \mp \f{\delta^i{}_j}{4\pi r} \;, \cr
  q^{\pm\,i} \ast \tilde{q}^{\pm}{}_j &= \tilde{q}^{\pm}{}_j \, q^{\pm\,i} \pm \f{\delta^i{}_j}{4\pi r} \;,
\end{align}
which descend directly from \eqref{EQ:1d_correlators_with_flavor}.

\subsection{OPE Algebra for One-Dimensional Flavor Currents}
The associative algebra introduced above has an important feature that can be deduced from the identities \eqref{EQ:star_product_qq}: the star product of any two operators has the structure
\begin{align}
  \mc{O}_1*\mc{O}_2 = \mc{O}_1\mc{O}_2 + \ldots
\end{align}
where the dots correspond to terms which are suppressed in inverse powers of $4\pi r$. In the $r\to\infty$ limit, the star product reduces to the usual commutative product.

As in \cite{Beem:2016cbd}, one can interpret $*$ as a deformation quantization of the chiral ring of the theory. For a detailed account on this the reader can refer to \cite{Beem:2016cbd,Etingof:2019guc}.

We now turn our attention to certain quadratic polynomials, $J^{\pm\,i}{}_j$, which are $\mr{SU}(N)$ Noether currents from the viewpoint of the one-dimensional theory, and show how these close a subalgebra on their own. In the context of superconformal gauge theories, the operators $J^{\pm\,i}{}_j$ are coupled to vector multiplet auxiliary fields and are referred to as ``moment map'' operators.

The one-dimensional action \eqref{EQ:1d_action_S2xS1} is invariant under the $\mr{SU}(N)$ flavor transformations 
\begin{align}\label{EQ:SU(Nf)_noether_variations}
  \delta q^{\pm\,i} &= - \mr{i}\vartheta^{\pm A} \, (\mr{t}_A)^i{}_j \, q^{\pm\,j} \;, \cr
  \delta\tilde{q}^{\pm}{}_{i} &= + \mr{i}\vartheta^{\pm A} \, \tilde{q}^{\pm}{}_{j} \, (\mr{t}_A)^j{}_i \;.
\end{align}
Here, $\vartheta^{\pm A}$ are constants and $(\mr{t}_A)^i{}_j$ are $N^2-1$ generators of $\mr{SU}(N)$. In our conventions, the generators $\mr{t}_A$ are normalized with
\begin{align}
  \tr(\mr{t}_A \mr{t}_B) &= \frac{1}{2} \delta_{AB} \;,
\end{align}
and satisfy the completeness relations
\begin{align}
  2\delta^{AB} (\mr{t}_A)^i{}_j (\mr{t}_B)^k{}_l &= \delta^i{}_l\,\delta^k{}_j - \frac{1}{N} \delta^i{}_j \, \delta^k{}_l \;.
\end{align}
The structure constants ${f_{AB}}^C$ are defined with
\begin{align}
  [\mr{t}_A, \mr{t}_B] &= \mr{i}f_{AB}{}^C\,\mr{t}_C \;,
\end{align}
while the Cartan-Killing form of $\mr{SU}(N)$ is defined with
\begin{align}
  g_{AB} &= f_{AC}{}^D f_{BD}{}^C \cr
  &= 2N\tr(\mr{t}_A \mr{t}_B) \cr
  &= N \delta_{AB} \;,
\end{align}

The Noether currents corresponding to this residual $\mr{SU}(N)$ flavor symmetry read
\begin{align}
  J^{\pm}{}_{\!A} &= \mp 8\pi N r \, \tilde{q}^{\pm}{}_i \, (\mr{t}_A)^i{}_j \, q^{\pm\,j} \;.
\end{align}
The normalization has been chosen for later convenience. If we convert the adjoint index $A$ to fundamental and anti-fundamental indices $i$, $j$, we find
\begin{align}\label{EQ:J_Noether}
  J^{\pm\,i}{}_j &= g^{AB} J^{\pm}{}_{\!A} \, (\mr{t}_B)^i{}_j \cr
  &= \mp 4\pi r \left(\tilde q^{\pm}{}_j \, q^{\pm\,i} - \frac{\delta^i{}_j}{N} \, \tilde{q}^{\pm}{}_k \, q^{\pm\,k}\right) \;.
\end{align}
All the components of these currents have vanishing one-point functions $\langle\!\langle J^{\pm\,i}{}_j \rangle\!\rangle = 0$.
These obey the star product identities
\begin{align}
  J^{\pm\,i}{}_j \ast q^{\pm\,k} &= J^{\pm\,i}{}_j\,q^{\pm\,k} + \delta^k{}_j\,q^{\pm\,i} - \f{1}{N}\,\delta^i{}_j\,q^{\pm\,k} \;, \cr
  q^{\pm\,k} \ast J^{\pm\,i}{}_j &= J^{\pm\,i}{}_j\,q^{\pm\,k} - \delta^k{}_j\,q^{\pm\,i} + \f{1}{N}\,\delta^i{}_j\,q^{\pm\,k} \;, \cr
  J^{\pm\,i}{}_j \ast \tilde{q}^{\pm}{}_k &= J^{\pm\,i}{}_j\,\tilde{q}^{\pm}{}_{k\,} - \delta^i{}_k\,\tilde{q}^{\pm}{}_j + \f{1}{N}\,\delta^i{}_j\,\tilde{q}^{\pm}{}_k \;, \cr
  \tilde{q}^{\pm}{}_k \ast J^{\pm\,i}{}_j &= J^{\pm\,i}{}_j\,\tilde{q}^{\pm}{}_{k\,} + \delta^i{}_k\,\tilde{q}^{\pm}{}_j - \f{1}{N}\,\delta^i{}_j\,\tilde{q}^{\pm}{}_k \;.
\end{align}
These show how $q^{\pm\,j}$ and $\tilde{q}^{\pm}{}_i$ indeed transform, respectively, in the fundamental and anti-fundamental representation of $\mr{SU}(N)$.
From the above identities one obtains
\begin{align}
  J^{\pm\,i}{}_j \ast J^{\pm\,k}{}_l &= J^{\pm\,i}{}_j\,J^{\pm\,k}{}_l + \delta^i{}_l\,J^{\pm\,k}{}_j - \delta^k{}_j\,J^{\pm\,i}{}_l - \left(\delta^k{}_j\delta^i{}_l-\f{1}{N}\delta^i{}_j\delta^k{}_l\right) \;.
\end{align}
The inverse powers of $4\pi r$ in the results above have been absorbed by carefully chosing the normalization in the definition \eqref{EQ:J_Noether}. Such powers can be reinstated by a simple rescaling of the currents and by counting the power of $J^{\pm}$ in each term.

\section{Surface Defects}\label{SEC:defects}
In this section we introduce supersymmetric surface defects on $S^2\times S^1$. In particular, we will consider $\mc{N}=(2,2)$ multiplets with support on $S^2$. Twisted chiral and anti-chiral multiplets, as well as vector multiplets, can be directly coupled to the bulk hypermultiplets through twisted superpotentials, thus producing an interacting theory of two- and three-dimensional quantum fields. Crucially, the localizing supercharge \eqref{EQ:localizing_supercharge} agrees with the standard choice of localizing supercharge for $\mc{N}=(2,2)$ theories in two dimensions \cite{Benini:2012ui, Doroud:2012xw}. This allows us to perform exact computations on the mixed-dimensional theory under consideration: the three-dimensional degrees of freedom can be localized as illustrated in previous sections, while the contributions coming from two-dimensional degrees of freedom can be addressed by using the approach originally presented in \cite{Gomis:2012wy}.

\subsection{Defect Subalgebra}\label{SEC:defect_subalgebra}
The $\widehat{\mf{su}}(1|1)$ superalgebra discussed in Section \ref{SEC:S2S1_localization} can be seen as a subalgebra of the $\mc{N}=(2,2)$ superalgebra on $S^2$, namely $\widehat{\mf{su}}(2|1)$. This fact can be used to insert into the path integral BPS surface defects wrapping $S^2$. Specifically, the full $\mf{su}(2|2)$ isometry superalgebra of $S^2 \times S^1$ reduces to the $\widehat{\mf{su}}(2|1)$ subalgebra by considering the linear combinations
\begin{align}
  \ms{Q} &= \tilde{\ms{Q}}_{\dot{1}} + \ms{Q}_{\dot{1}} \;, &
  \ms{S} &= \ms{S}_{\dot{2}} + \tilde{\ms{S}}_{\dot{2}} \;, \cr
  \tilde{\ms{Q}} &= \tilde{\ms{Q}}_{\dot{2}} - \ms{Q}_{\dot{2}} \;, &
  \tilde{\ms{S}} &= \ms{S}_{\dot{1}} - \tilde{\ms{S}}_{\dot{1}} \;.
\end{align}
Their non-trivial anticommutation relations are
\begin{align}
  \{\ms{Q}, \tilde{\ms{Q}}\} &= 2r^{-1}\ms{J}_+ \cr
  \{\ms{Q}, \ms{S}\} &= 2\mr{i}r^{-1} (\ms{J}_{3\vphantom{\dot{1}}} - \ms{R}_{\dot{1}\dot{2}}) \;, \cr
  \{\ms{S}, \tilde{\ms{S}}\} &= 2r^{-1}\ms{J}_- \;, \cr
  \{\tilde{\ms{Q}}, \tilde{\ms{S}}\} &= 2\mr{i}r^{-1} (\ms{J}_{3\vphantom{\dot{1}}} + \ms{R}_{\dot{1}\dot{2}}) \;,
\end{align}
which are the anticommutation relations of the $\widehat{\mf{su}}(2|1)$ superalgebra. The latter has four supercharges and its bosonic part is $\mf{su}(2) \oplus \mf{u}(1)$, where $\mf{su}(2)$ is the isometry algebra of $S^2$ spanned by $(\ms{J}_\pm, \ms{J}_3)$ and $\mf{u}(1)$ is the vector-like R-symmetry algebra spanned by $\ms{R}_{\dot{1}\dot{2}}$. In particular, the $\mf{u}(1)$ R-symmetry of the defect theory on $S^2$ coincides with the Cartan subalgebra of the $\mf{su}(2)_{\mr{C}}$ R-symmetry algebra of the three-dimensional theory on $S^2 \times S^1$.

The eight-parameter family \eqref{EQ:killing_solutions} of Killing spinors $\xi^{a\dot{a}}$ on $S^2 \times S^1$ reduces to the four-parameter family of Killing spinors $\epsilon$, $\tilde{\epsilon}$ via the map 
\begin{align}\label{EQ:killing_spinors_dim_reduction}
  \xi^{1\dot{1}} &= \tilde{\epsilon} \;, &
  \xi^{2\dot{1}} &= + \gamma^{\ms{3}} \tilde{\epsilon} \;, \cr
  \xi^{2\dot{2}} &= \epsilon \;, &
  \xi^{1\dot{2}} &= - \gamma^{\ms{3}} \epsilon \;.
\end{align}
where $\epsilon$, $\tilde{\epsilon}$ satisfy the Killing spinor equations on $S^2$,
\begin{align}
  \nabla_{\underline{\mu}}\epsilon &= \f{1}{2r} \gamma_{\underline{\mu}} \gamma^3 \epsilon \;, &
  \nabla_{\underline{\mu}}\tilde{\epsilon} &= -\f{1}{2r} \gamma_{\underline{\mu}} \gamma^3 \tilde{\epsilon} \;.
\end{align}
Here, the underlined index $\underline{\mu}$ runs over the coordinates $\theta,\varphi$ of the $S^2$ defect manifold.
In terms of the parameters $\xi_0^{a\dot{a}}$, the map \eqref{EQ:killing_spinors_dim_reduction} yields the identifications 
\begin{align}\label{EQ:killing_spinors_dim_reduction_param}
  \xi_0^{2\dot{1}} &= -\gamma^{\ms{3}}\xi_0^{1\dot{1}} \;, & \xi_0^{1\dot{2}} &= +\gamma^{\ms{3}}\xi_0^{2\dot{2}} \;.
\end{align}
The Killing spinor $\epsilon$ depends on $\xi_0^{2\dot{2}}$ and the variation $\delta_\epsilon$ encodes the action of the supercharges $\tilde{\ms{Q}}$ and $\ms{S}$. Correspondingly, the Killing spinor $\tilde{\epsilon}$ depends on $\xi_0^{1\dot{1}}$ and the variation $\delta_{\tilde{\epsilon}}$ encodes the action of the supercharges $\ms{Q}$ and $\tilde{\ms{S}}$. Indeed,
\begin{align} 
\delta_\xi\mc{O} &= \delta_{\tilde{\epsilon}} \mc{O} + \delta_\epsilon \mc{O} \;.
\end{align}

\subsection{Defect Multiplets}
We will now find the defect multiplets descending from the reduction in (\ref{EQ:killing_spinors_dim_reduction}). These are twisted chiral (and anti-chiral) multiplets on $S^2$.
The supersymmetry variations of a two-dimensional twisted chiral multiplet formally mix objects transforming in conjugate representations. This makes sense if the twisted chiral multiplet has a flavor\footnote{The flavor symmetry might, or might not, be gauged} group admitting a pseudoreal representation. This occurrence suits a twisted chiral multiplet descending from a higher dimensional hypermultiplet.

In the case under consideration, already discussed in Section \ref{SEC:correlators}, where the residual flavor group is $\mr{SU}(N)$, we use the embedding of the $\v{N}\oplus\overline{\v{N}}$ representation of $\mr{SU}(N)$ into the pseudoreal fundamental representation of the original flavor group $\mr{USp}(2N)$. This allows us to collect the components $(q_a, \tilde{q}_a, \psi_{\dot{a}}, \tilde{\psi}_{\dot{a}}, G_a, \tilde{G}_a)$ into the $\mr{USp}(2N)$ multiplets $(Q_a, \Psi_{\dot{a}}, \mc{G}_a)$ with
\begin{align}
  Q_a{}^I &= (q_a{}^i, \tilde{q}_{a\,i}) \;, \cr
  \Psi_{\dot{a}}{}^I &= (\psi_{\dot{a}}{}^i, \tilde{\psi}_{\dot{a}\,i}) \;, \cr
  \mc{G}_a{}^I &= (G_a{}^i, \tilde{G}_{a\,i}) \;,  
\end{align}
where
$I = 1, \ldots, 2N$ are $\mr{USp}(2N)$ indices. The reality conditions \eqref{EQ:hyper_reality_conditions} translate to
\begin{align}
  (Q^{a\,I})^* &= \Omega_{IJ} \, Q_a{}^J \;, \cr
  (G^{a\,I})^* &= \Omega_{IJ} \, G_a{}^{J} \;,
\end{align}
with
\begin{align}
  \Omega_{IJ} &= \begin{pmatrix} 0 && \v{1}_N \\ - \v{1}_N && 0 \end{pmatrix}
\end{align}
being a $\mr{USp}(2N)$-invariant form.
The supersymmetry transformations of the new fields can be readily derived from \eqref{EQ:susy_scalars} and \eqref{EQ:susy_fermions_off_shell} and read
\begin{align}
  \delta Q^{a} &= \xi^{a\dot{a}} \Psi_{\dot{a}} \;, \cr
  \delta \Psi_{\dot{a}} &= \mr{i} \gamma^\mu \xi_{a\dot{a}} \mr{D}_\mu Q^a + \mr{i} \eta_{a\dot{a}} Q^a - \mr{i} \xi_{a\dot{c}} {\Phi_{\dot{a}}}^{\dot{c}} Q^a - \mr{i} \nu_{a\dot{a}} \mc{G}^a \;, \cr
  \delta \mc G^a &=  \nu^{a\dot{a}} \gamma^\mu \mr{D}_\mu \Psi_{\dot{a}} + \nu^{a\dot{a}} {\Phi_{\dot{a}}}^{\dot{c}} \Psi_{\dot{c}} \;. 
\end{align}

Let us now consider the reduction to two dimensions of a three-dimensional hypermultiplet. For the moment, we omit the coupling with a background vector multiplet. We make the following identifications:
\begin{align}
  Y &= (Q^1 + Q^2)/2 \;, \cr
  \tilde{Y} &= (Q^1 - Q^2)/2 \;, \cr
  \chi &= \Psi_{\dot{1}} \;, \cr
  \tilde{\chi} &= \Psi_{\dot{2}} \;, \cr
  F &= (\mc{G}^1 - \mc{G}^2)/2 + \partial_{\ms{3}} (Q^1 - Q^2)/2 \;, \cr
  \tilde{F} &= (\mc{G}^1 + \mc{G}^2)/2 - \partial_{\ms{3}} (Q^1 + Q^2)/2 \;.
\end{align}
These can be interpreted as the components of a pair of two-dimensional twisted chiral and anti-chiral multiplets $(Y, \mr{P}_-\chi, \mr{P}_+\tilde{\chi}, F)$ and $(\tilde{Y}, \mr{P}_+\chi, \mr{P}_-\tilde{\chi}, \tilde{F})$. With $\mr{P}_\pm$ we denote the chiral projectors 
\begin{align}
  \mr{P}_\pm &= \f{1\pm\gamma^{\ms{3}}}{2} \;.
\end{align}
If the twisted multiplets have Weyl weight $\Delta$, their field components transform under supersymmetry with
\begin{align}\label{EQ:twisted_chiral_susy_variations}
  \delta Y &= \tilde{\epsilon}\mr{P}_-\chi + \epsilon\mr{P}_+\tilde{\chi} \;, \cr
  \delta \tilde Y &= \tilde{\epsilon}\mr{P}_+\chi - \epsilon\mr{P}_-\tilde{\chi} \;, \cr
  \delta \chi &= 2\mr{i} [\gamma^{\underline{\mu}}\mr{P}_+\epsilon\,\mr{D}_{\underline{\mu}}Y + \gamma^{\underline{\mu}}\mr{P}_-\epsilon\,\mr{D}_{\underline{\mu}}\tilde{Y} - \mr{P}_-\epsilon(F+\Delta Y/r) - \mr{P}_+\epsilon(\tilde{F}+\Delta\tilde{Y}/r)] \;, \cr
  \delta \tilde{\chi} &= 2\mr{i}[\gamma^{\underline{\mu}} \mr{P}_-\tilde{\epsilon}\,\mr{D}_{\underline{\mu}}Y - \gamma^{\underline{\mu}}\mr{P}_+\tilde{\epsilon}\,\mr{D}_{\underline{\mu}}\tilde{Y} - \mr{P}_+\tilde{\epsilon} (F + \Delta Y/r) + \mr{P}_-\tilde{\epsilon} (\tilde{F} + \Delta\tilde{Y}/r)] \;, \cr
  \delta F &= -\tilde{\epsilon}\gamma^{\underline{\mu}}\mr{D}_{\underline{\mu}}\mr{P}_-\chi - \epsilon\gamma^{\underline{\mu}}\mr{D}_{\underline{\mu}}\mr{P}_+\tilde{\chi} - \Delta\tilde\epsilon\mr{P}_-\chi/r - \Delta\epsilon\mr{P}_+\tilde{\chi}/r \;, \cr
  \delta \tilde{F} &= - \tilde{\epsilon}\gamma^{\underline{\mu}}\mr{D}_{\underline{\mu}}\mr{P}_+\chi + \epsilon\gamma^{\underline{\mu}}\mr{D}_{\underline{\mu}}\mr{P}_-\tilde{\chi} - \Delta\tilde{\epsilon}\mr{P}_+\chi/r + \Delta\epsilon\mr{P}_-\tilde{\chi}/r \;.
\end{align}
In fact, by using the identifications \eqref{EQ:killing_spinors_dim_reduction}, the variations above hold for $\Delta = 0$.

For a hypermultiplet coupled to a background vector multiplet, one finds
\begin{align}
F &\mapsto F - \mr{i} (A_{\ms{3}} + \mr{i}\Phi_{\dot{1}\dot{2}}) \tilde{Y} \;, \cr
\tilde F &\mapsto \tilde{F} + \mr{i} (A_{\ms{3}} - \mr{i}\Phi_{\dot{1}\dot{2}}) Y \;.
\end{align}
At the same time, one needs to modify the supersymmetric variations with
\begin{align}
\delta\chi &\mapsto \delta\chi + 2\mr{i}(\mr{P}_-\tilde{\epsilon}\,\Phi_{\dot{2}\dot{2}}Y - \mr{P}_+\tilde{\epsilon}\,\Phi_{\dot{2}\dot{2}}\tilde{Y}) \;, \cr
\delta\tilde{\chi} &\mapsto \delta\tilde{\chi} - 2\mr{i}(\mr{P}_+\epsilon\,\Phi_{\dot{1}\dot{1}}Y + \mr{P}_-\epsilon\,\Phi_{\dot{1}\dot{1}}\tilde{Y}) \;, \cr
\delta F &\mapsto \delta F + \Phi_{\dot{2}\dot{2}}\,\tilde{\epsilon}\mr{P}_+\tilde{\chi} - \Phi_{\dot{1}\dot{1}}\,\epsilon\mr{P}_-\chi \;, \cr
\delta \tilde{F} &\mapsto \delta\tilde{F} + \Phi_{\dot{2}\dot{2}}\,\tilde{\epsilon}\mr{P}_-\tilde{\chi} + \Phi_{\dot{1}\dot{1}}\,\epsilon\mr{P}_+\chi \;.
\end{align}
However, since on our background vector multiplet $\Phi_{\dot{1}\dot{1}} = \Phi_{\dot{2}\dot{2}} = 0$ all these modifications do not affect the form of \eqref{EQ:twisted_chiral_susy_variations}. This agrees with the fact that twisted chiral multiplets cannot be minimally coupled to vector multiplets \cite{Gomis:2012wy}.
With \eqref{EQ:twisted_chiral_susy_variations} we can couple the twisted chiral multiplet reduced from three dimensions, which has $\Delta = 0$, to general twisted chiral multiplets of arbitrary Weyl weight $\Delta_i$.

\subsection{Coupled System and Localization}
We can now couple the defect multiplets to the three-dimensional theory via twisted superpotentials $W = W(Y^i)$ that depend holomorphically on a set of complex scalars $Y^i$. The superscript $i$ labels the $i$-th multiplet and can be regarded as a multi-index also encoding flavor indices.
The twisted superpotential term
\begin{align}\label{EQ:twisted_superpotential}
S_W = \int_{S^2} \star_{S^2}\bigg(\f{\mr{i}}{r}W - \mr{i}\bigg(F^i + \f{\Delta_i}{r}Y^i\bigg) \partial_i W - \f{1}{2} \partial_i \partial_j W \, \tilde{\chi}^i_+ \chi^j_- \bigg)
\end{align}
and its twisted anti-holomorphic counterpart
\begin{align}\label{EQ:twisted_superpotential_tilde}
S_{\tilde{W}} = \int_{S^2} \star_{S^2}\bigg(\f{\mr{i}}{r} \tilde{W} + \mr{i} \bigg(\tilde{F}^i  + \f{\Delta_i}{r} \tilde{Y}^i\bigg) \partial_i \tilde{W} - \f{1}{2} \partial_i \partial_j \tilde{W} \, \tilde{\chi}^i_- \chi^j_+\bigg)  
\end{align}
are both supersymmetric.

Twisted chiral multiplets coupled to twisted superpotentials were localized in \cite{Gomis:2012wy}. The outcome is that the only non-trivial contributions are due to $W(Y^i)/r$ and $\tilde{W}(\tilde{Y}^i)/r$ evaluated at the poles of $S^2$. These appear as local insertions in the path integral of the full theory. Specifically, the BPS locus for the two-dimensional degrees of freedom is \cite{Gomis:2012wy,Pan:2017zie}
\begin{align}
Y^{(\mr{2d})} &= Y^{(\mr{2d})}_0 \;, & F^{(\mr{2d})} &= -\Delta r^{-1} \, Y^{(\mr{2d})}_0 \;, \cr
\tilde{Y}^{(\mr{2d})} &= \tilde{Y}^{(\mr{2d})}_0 \;, & \tilde{F}^{(\mr{2d})} &= -\Delta r^{-1} \, \tilde{Y}^{(\mr{2d})}_0 \;, 
\end{align}
where $Y^{(\mr{2d})}_0$ and $\tilde{Y}^{(\mr{2d})}_0$ are constant values on $S^2$. On the other hand, by direct computation we find that the BPS locus for the twisted chiral multiplet descending from the three-dimensional hypermultiplet fulfills
\begin{align}
  \partial_\varphi Y^{(\mr{3d})} &= 0 \;, &
  F^{(\mr{3d})} &= r^{-1} \cot(\theta/2) \, \partial_\theta Y^{(\mr{3d})} \;, \cr
  \partial_\varphi \tilde{Y}^{(\mr{3d})} &= 0 \;, &
  \tilde{F}^{(\mr{3d})} &= r^{-1} \tan(\theta/2) \, \partial_\theta \tilde{Y}^{(\mr{3d})} \;.
\end{align}
As a result, when evaluated on the BPS solutions, the holomorphic and the anti-holomorphic twisted superpotential actions give
\begin{align}
  S_W|_{\mr{BPS}}
  &= 4\pi\mr{i}r \, W(Y^{(\mr{2d})}_0, Y^{(\mr{3d})}_{\mr{N}}) \;, \cr
  S_{\tilde W}|_{\mr{BPS}}
  &= 4\pi\mr{i}r \, \tilde{W}(\tilde{Y}^{(\mr{2d})}_0, \tilde{Y}^{(\mr{3d})}_{\mr{S}}) \;. 
\end{align}
These contributions are localized at the two points of intersection between $\mh{M}_v$ and the $S^2$ where the defect multiplets have their support, namely $\theta=0$, $t=0$ and $\theta=\pi$, $t=0$. When evaluated on such points, $Y^{(\mr{3d})}$ and $\tilde{Y}^{(\mr{3d})}$ can be written, respectively, in terms of the BPS operators $q^+(0)$, $\tilde{q}^+(0)$ and $q^-(0)$, $\tilde{q}^-(0)$ that appear in the one-dimensional action \eqref{EQ:1d_action_S2xS1}.
Plugged into the path integral, the contributions of twisted superpotentials correspond to insertions of exponential operators $\mc{V}$ and $\tilde{\mc{V}}$ at the north and south pole of $S^2$ respectively. Specifically, 
\begin{align}
  \mc{V} &= e^{-4\pi\mr{i}r W(Y^{(\mr{2d})}_0,\,q^+(0),\,\tilde{q}^+(0))} \;, \cr
  \tilde{\mc{V}} &= e^{- 4\pi\mr{i}r \tilde{W}(\tilde{Y}^{(\mr{2d})}_0,\,q^-(0),\,\tilde{q}^-(0))} \;.
\end{align}

The action of a two-dimensional $\mc{N}=(2,2)$ vector multiplet $(A,\lambda,\tilde{\lambda},\sigma_1,\sigma_2,D)$ 
can effectively be regarded as that of an adjoint twisted chiral multiplet, known also as ``field-strenght multiplet'', with components
\begin{align}
  (\sigma_2+\mr{i}\sigma_1,\lambda,\tilde{\lambda},D-\sigma_2/r+\mr{i}f_{\ms{1}\ms{2}}) \;.
\end{align}
This means that the superpotential interaction discussed above can also be used to couple bulk hypermultiplets to vector multiplets on $S^2$.
The BPS locus for a vector multiplet is
\begin{align}
  \sigma_1 &= -\f{\mf{m}}{2r} \;, &
  \sigma_2 &= \sigma_{2,0} \;, \cr
  f_{\ms{1}\ms{2}} &= +\f{\mf{m}}{2r^2} \;, &
  D &= 0 \;,
\end{align}
where $\sigma_{2,0}$ is constant and $\mf{m}$ is the monopole charge associated with the field strength $f_{\ms{1}\ms{2}}$.

More generally, since, as mentioned earlier, the localizing supercharge \eqref{EQ:localizing_supercharge} employed in Section \ref{SEC:S2S1_localization} is compatible with the standard choice of localizing supercharge adopted in \cite{Benini:2012ui, Doroud:2012xw, Gomis:2012wy}, we are free to introduce also $\mc{N}=(2,2)$ chiral and anti-chiral multiplets on $S^2$, minimally coupled to the two-dimensional vector multiplet.
Schematically, we can write the full partition function in the presence of the defect as
\begin{align}
Z_{\mr{3d/2d}} = \sum_{\mf{m}} \int\mr{d}\mc{Y}\;Z_{\text{1-loop}}(\mf{m},\mc{Y}) \int [\mr{d}\tilde{q}^+][\mr{d}q^+][\mr{d}\tilde{q}^-][\mr{d}q^-] \; e^{-S_{\text{1d}}(\tilde{q}^\pm,q^\pm)} \; \mc{V}\,\tilde{\mc{V}}
\end{align}
where with $\mf{m}$ and $\mc{Y}$ we collectively denote the discrete and continuous part of the two-dimensional BPS locus, while $Z_{\text{1-loop}}$ denotes the one-loop determinant associated with the two-dimensional multiplets. Notice how the only action terms that appear in the above are $S_{\text{1d}}$ and the superpotential terms $\mc{V}$ and $\tilde{\mc{V}}$. This is due to the fact that all the action terms for an $\mc{N}=(2,2)$ theory are $\v{Q}$-exact, with the exception of \eqref{EQ:twisted_superpotential} and \eqref{EQ:twisted_superpotential_tilde} \cite{Benini:2012ui,Gomis:2012wy}. Supersymmetric localization has turned the path integral associated with a theory of three- and two-dimensional degrees of freedom into the path integral of coupled zero- and one-dimensional quantum systems.

Finally, one can compute correlators by adding insertions of BPS operators coming either from the $\mc{N}=(2,2)$ multiplets (e.g.\ \cite{Hosomichi:2017dbc, Panerai:2018ryw}) or from the bulk theory. The latter have been extensively discussed in Section \ref{SEC:correlators}, whether the former appear in the quantum mechanics as insertions of local operators at the intersection point, namely $t=0$.

\section*{Acknowledgements}
We thank Joseph A. Minahan for reading the manuscript and providing suggestions. We also thank Lorenzo Ruggeri for collaboration on related topics. The work of R.P.\ is supported by the Knut and Alice Wallenberg Foundation under grant Dnr KAW 2015.0083. The work of A.P.\ is supported by the ERC STG Grant 639220. The work of K.P.\ is supported by the grant “Geometry and Physics” from the Knut and Alice Wallenberg foundation.

\newpage
\appendix

\section{Conventions and Notation}
We denote spacetime indices with Greek letters $\mu$, $\nu$, \ldots and $\mr{SO}(3)$ frame indices with a sans serif font $\ms{a}$, $\ms{b}$, \ldots.
Spinor indices are denoted with Greek letters $\alpha$, $\beta$, \ldots. These are raised and lowered from the left with the $\mr{SU}(2)$ invariant tensor $\epsilon^{12} = - \epsilon_{12} = 1$. The same applies for $\mr{SU}(2)_{\mr{H}}$ and $\mr{SU}(2)_{\mr{C}}$ indices, which are labelled by undotted $a$, $b$, \ldots and dotted $\dot{a}$, $\dot{b}$, \ldots Latin letters.

Let $\xi,\psi$ and $\eta$ be either commuting or anti-commuting spinors. The spinor contraction is defined as
\begin{equation}
	\xi\psi = \xi^\alpha \psi_\alpha \;,
\end{equation}
and the following Fierz identity holds:
\begin{equation}
	\xi_\alpha(\psi\eta) + (\xi\psi) \eta_\alpha + \xi_\beta\psi_\alpha\eta^\beta = 0 \;.
\end{equation}

The ``flat'' gamma matrices are defined as the Pauli matrices $\gamma^{\ms{a}}=\sigma^{\ms{a}}$ and the ``curved'' ones are given by contraction with the dreibein as $\gamma^\mu = e_{\ms{a}}^\mu\gamma^{\ms{a}}$. These satisfy
\begin{align}
	\gamma_{\mu}\gamma_{\nu} &= g_{\mu\nu} + \mr{i} \epsilon_{\mu\nu\rho} \gamma^\rho \;, \\
	(\gamma^\mu)_\alpha{}^\beta (\gamma_{\mu})_\gamma{}^\delta &= 2 \delta_{\alpha}{}^\delta \delta_{\gamma}{}^\beta -\delta_{\alpha}{}^\beta \delta_{\gamma}{}^\delta\,.
\end{align}

Let us consider a three-dimensional Riemannian manifold $\mh{M}$ with metric $g$. We define the musical isomorphism $\flat$ as the map from a vector $v$ to a one-form $v^\flat$ such that, for any vector $w$,
\begin{align}
  \iota_w v^\flat = g(v,w) \;.
\end{align}
The Lie derivative along a vector field $v$ acting on an $r$-form $\omega$ is defined as
\begin{equation}
	\mathcal{L}_v\omega = (\mr{d}\iota_v + \iota_v\mr{d})\omega \;.
\end{equation}
When acting on a spinor $\psi$, the Lie derivative reads
\begin{align}
  \mc{L}_v \psi = v^\mu \nabla_\mu \psi + \f{\mr{i}}{2}\epsilon^{\mu\nu\rho}\nabla_\mu v_\nu \gamma_\rho \psi \;.
\end{align}
With $\star$, we denote the Hodge dual, a linear map from an $r$-form to a $(3-r)$-form such that, for any two $r$-forms $\alpha$, $\beta$,
\begin{align}
  \alpha \wedge \star\beta = g(\alpha,\beta) \, e^\ms{1} \wedge e^\ms{2} \wedge e^\ms{3} \;.
\end{align}

In our conventions, the Lichnerowicz--Weitzenb\"ock formula reads
\begin{align}
  \slashed{\nabla}^2\psi = \bigg(\nabla^2 - \f{R}{4}\bigg)\psi \;.
\end{align}
This fixes the form of the auxiliary equation for a conformal Killing spinor $\xi^{\alpha\dot{\alpha}}$,
\begin{align}
  \bigg(\mr{D}^2 + \f{R}{8}\bigg)\xi^{\alpha\dot{\alpha}} = 0 \;.
\end{align}

\section{Superalgebra Closure}\label{APP:algebra_closure}
By imposing the Killing spinor equation \eqref{EQ:killing_spinor_equation} one finds that the supersymmetry variations \eqref{EQ:susy_scalars} and \eqref{EQ:susy_fermions} close on the $\mc{N}=4$ superalgebra associated with $\mh{M}$. Let us consider variations generated by Killing spinors $\xi_1^{a\dot{a}}$ and $\xi_2^{a\dot{a}}$. Their action on fields is given by
\begin{align}
  \{\delta_1,\delta_2\}q^a &= \mc{L}_vq^a + (R^a{}_c - \mr{i}\iota_v(A_{\mr{H}})^{a}{}_c)q^c + 1/2\,\rho q^a + \mr{i}\Xi q^a \;, \cr
  \{\delta_1,\delta_2\}\tilde{q}^a &= \mc{L}_v\tilde{q}^a + (R^a{}_c - \mr{i}\iota_v(A_{\mr{H}})^{a}{}_c)\tilde{q}^c + 1/2\,\rho\tilde{q}^a - \mr{i}\tilde{q}^a\Xi \;,
\end{align}
and
\begin{align}\label{EQ:closure_fermions}
  \{\delta_1,\delta_2\}\psi_{\dot{a}} &= \mc{L}_v\psi_{\dot{a}} - (\bar{R}_{\dot{a}}{}^{\dot{c}}\psi_{\dot{c}} + \mr{i}\iota_v(A_{\mr{C}})_{\dot{a}}{}^{\dot{c}})\psi_{\dot{a}} + \rho\psi_{\dot{a}} + \mr{i}\Xi\psi_{\dot{a}} + \mc{E}_{\dot{a}} \;, \cr
  \{\delta_1,\delta_2\}\tilde{\psi}_{\dot{a}} &= \mc{L}_v\tilde{\psi}_{\dot{a}} - (\bar{R}_{\dot{a}}{}^{\dot{c}}\tilde{\psi}_{\dot{c}} + \mr{i}\iota_v(A_{\mr{C}})_{\dot{a}}{}^{\dot{c}})\tilde{\psi}_{\dot{a}} + \rho\psi_{\dot{a}} - \mr{i}\tilde{\psi}_{\dot{a}}\Xi + \tilde{\mc{E}}_{\dot{a}} \;.
\end{align}
On the r.h.s.\ of the above, we find the generators of various bosonic symmetries. These are
\begin{itemize}
\item
conformal isometries generated by the conformal Killing vector
\begin{align}
  v^{\mu} = \mr{i}\xi_1^{u\dot{u}}\gamma^\mu\xi_{2,u\dot{u}}^{\phantom{\dot{u}}} \;;
\end{align}
\item
$\mf{su}(2)_{\mr{H}}\oplus\mf{su}(2)_{\mr{C}}$ R-symmetry transformations that act on undotted and dotted indices with
\begin{align}
  R_{ac} &= \mr{i}({\xi_{1,(a}}^{\dot{u}}\eta_{2,c)\dot{u}} + {\xi_{2,(a}}^{\dot{u}}\eta_{1,c)\dot{u}}) \;, \cr
  \bar{R}_{\dot{a}\dot{c}} &= \mr{i}({{\xi_1}^u}\vphantom{\xi}_{(\dot{a}}\eta_{2,u\dot{c})} + {{\xi_2}^u}\vphantom{\xi}_{(\dot{a}}\eta_{1,u\dot{c})}) \;,
\end{align}
and through the background connections $(A_{\mr{H}})^a{}_c$ and $(A_{\mr{C}})_{\dot{a}}{}^{\dot{c}}$;
\item
dilatations, acting on fields with $\Delta\rho$, where
\begin{align}
  \rho = \mr{i}(\xi_1^{u\dot{u}}\eta_{2,u\dot{u}}^{\phantom{\dot{u}}} + \xi_2^{u\dot{u}}\eta_{1,u\dot{u}}^{\phantom{\dot{u}}})
\end{align}
is the dilatation parameter and $\Delta$ is the Weyl weight of the field;
\item
background gauge transformations with gauge parameter
\begin{align}
  \Xi = {{\xi_1}^u}\vphantom{\xi}_{(\dot{a}}\xi_{2,u\dot{c})}\Phi^{\dot{a}\dot{c}} - \iota_{v_{12}}A \;.
\end{align}
\end{itemize}

The algebra closes on shell on the fermionic components of the hypermultiplet. In fact, in \eqref{EQ:closure_fermions}, two additional terms appear, $\mc{E}_{\dot{a}}$ and $\tilde{\mc{E}}_{\dot{a}}$, that vanish on the solutions of the equations of motion for $\psi_{\dot{a}}$ and $\tilde{\psi}_{\dot{a}}$.
These terms can be eliminated for a certain choice of Killing spinors $\xi^{a\dot{a}}$, by the introduction of auxiliary fields $G^a$ and $\tilde{G}^a$ and by an appropriate modification of the supersymmetry variations, as described in Section \ref{SEC:setup}.

\section{Details of the Cohomological Approach}\label{APP:1d_action}
In this Appendix we outline the computation that leads to the cohomological construction illustrated in Section \ref{SEC:localization}.

On the solutions of \eqref{EQ:hyper_bps} and through the identities in \eqref{EQ:auxiliary_identities}, one can rewrite the auxiliary term in the off-shell Lagrangian $\mh{L}$ as
\begin{align}
  \tilde{G}^aG_a &= 4|v|^{-2}\bar{X}^{\dot{a}\dot{c}}\,\tilde{\Upsilon}_{\dot{a}}\Upsilon_{\dot{c}} \;.
\end{align}
where
\begin{align}
  \Upsilon_{\dot{a}} &= \gamma^\mu\xi_{a\dot{a}}\mr{D}_\mu q^a + \eta_{a\dot{a}}q^a - \xi_{a\dot{c}}{\Phi_{\dot{a}}}^{\dot{c}}q^a \;, \cr
  \tilde{\Upsilon}_{\dot{a}} &= \gamma^\mu\xi_{a\dot{a}}\mr{D}_\mu\tilde{q}^a + \tilde{q}^a\eta_{a\dot{a}} + \xi_{a\dot{c}}\tilde{q}^a{\Phi_{\dot{a}}}^{\dot{c}} \;.
\end{align}
After some manipulation, one finds
\begin{align}
  \tilde{G}^aG_a &= -\mh{L} + \operatorname{div}j \;,
\end{align}
up to terms that cancel on the solutions of \eqref{EQ:vector_bps} and when setting $\psi_{\dot{a}}=\tilde{\psi}_{\dot{a}}=0$. Here we have introduced the vector field
\begin{align}
  j^\mu = |v|^{-2}\Big(&\epsilon^{\mu\nu\rho}v_\rho X_{ac}(\tilde{q}^a\mr{D}_\nu q^c - \mr{D}_\nu\tilde{q}^aq^c) - 1/4\,\mr{D}^{\mu}|v|^2 \tilde{q}^aq_a + 4\bar{X}^{\dot{a}\dot{c}}\xi_{(a\dot{a}}\gamma^\mu\xi_{c)\dot{e}}\tilde{q}^a\Phi^{\dot{e}}\vphantom{\Phi}_{\dot{c}}q^c\Big) \;.
\end{align}
Now, $j$ is singular on $\mh{M}_v$, but we assume it to be well defined on $\mh{M}\setminus\mh{M}_v$.

Then, we consider
\begin{align}
  \iota_v(\star\mh{L}) = \iota_v\mr{d}(\star j^\flat) = (\mc{L}_v-\mr{d}\iota_v)(\star j^\flat)
\end{align}
and notice that if
\begin{align}\label{EQ:lie_d_identity}
  \mc{L}_v(\star j^\flat) = 0
\end{align}
one can immediately conclude that
\begin{align}
  \mr{d}_v(\star\mh{L} + \iota_v{\star j^\flat}) = 0 \;,
\end{align}
where $\alpha_1 = \iota_v{\star j^\flat}$ is now globally defined on $\mh{M}$.
The identity \eqref{EQ:lie_d_identity} can be proven by enforcing
\begin{align}\label{EQ:vector_closure}
  \iota_vF &= -\operatorname{D}(\bar{X}_{\dot{u}\dot{v}}\Phi^{\dot{u}\dot{v}}) \;, \cr
  \iota_v\operatorname{D}\Phi_{\dot{a}\dot{c}} &= -4\mr{i}\xi_{u(\dot{a}}\eta^{u\dot{u}}\Phi_{\dot{c})\dot{u}} + \mr{i}\bar{X}_{\dot{u}(\dot{a}}[\Phi_{\dot{c})\dot{v}},\Phi^{\dot{u}\dot{v}}] \;,
\end{align}
and
\begin{align}\label{EQ:hyper_closure}
  \iota_v\operatorname{D}q^a &= -2\mr{i}\xi^{a\dot{a}}\eta_{c\dot{a}}q^c - \mr{i}\bar{X}_{\dot{a}\dot{c}}\Phi^{\dot{a}\dot{c}}q^a \;, \cr
  \iota_v\operatorname{D}\tilde{q}^a &= -2\mr{i}\xi^{a\dot{a}}\eta_{c\dot{a}}\tilde{q}^c + \mr{i}\bar{X}_{\dot{a}\dot{c}}\tilde{q}^a\Phi^{\dot{a}\dot{c}} \;,
\end{align}
which come from \eqref{EQ:vector_bps} and \eqref{EQ:hyper_bps}.

\section{BPS Loci}
In this appendix we look at the problem of finding the explicit solutions of the BPS equation for both the background vector multiplet and the hypermultiplet.

\subsection{Background Vector Multiplet}\label{APP:background_gauge}
We start by considering the BPS equations \eqref{EQ:vector_bps} for the vector multiplet. We assume that all solutions are in the Cartan subalgebra of $\mf{g}$.

Combining \eqref{EQ:vector_closure} with the reality conditions \eqref{EQ:vector_reality_conditions} gives
\begin{align}
  \Phi_{\dot{1}\dot{1}} &= e^{-\mr{i}\varphi} \Phi_{0,\dot{1}\dot{1}}(\theta,t) \;, \\
  \Phi_{\dot{1}\dot{2}} &= \Phi_{\dot{1}\dot{2}}(\theta,t) \;, \\
  \Phi_{\dot{2}\dot{2}} &= e^{+\mr{i}\varphi} \Phi_{0,\dot{2}\dot{2}}(\theta,t) \;,
\end{align}
and
\begin{align}
  F_{\varphi\theta} &= 0 \;, & \partial_\theta(\sin\theta(\Phi_{0,\dot{1}\dot{1}} - \Phi_{0,\dot{2}\dot{2}})) &= 0 \;, \cr
  F_{\varphi t} &= 0 \;, & \partial_t(\Phi_{0,\dot{1}\dot{1}} - \Phi_{0,\dot{2}\dot{2}}) &= 0 \;.
\end{align}
If we impose regularity at the poles, we find that
\begin{align}
  \Phi_{0,\dot{1}\dot{1}} = \Phi_{0,\dot{2}\dot{2}} \;.
\end{align}

We can now plug these ansatze in \eqref{EQ:vector_background_sols} and find, still taking into account the reality conditions on the fields,
\begin{align}
  F_{t\theta} &= 0 \,, \cr
  D_{12} &= 0 \;, \cr
  D_{11} &= D_{22} \,,
\end{align}
together with
\begin{align}
  \partial_t\Phi_{0,\dot{1}\dot{1}} + \mr{i}\beta\,\partial_\theta\Phi_{\dot{1}\dot{2}} &= 0 \;, \cr
  - \mr{i}\sin\theta\,\partial_t\Phi_{\dot{1}\dot{2}} + \beta\partial_\theta(\sin\theta\,\Phi_{0,\dot{1}\dot{1}}) &= 0 \;.
\end{align}
The configuration where $\Phi_{0,\dot{1}\dot{1}}$ vanishes and $\Phi_{\dot{1}\dot{2}}$ is a real constant is the only solution of the above that is regular everywhere.

\subsection{Hypermultiplet}\label{APP:BPS_hyper}
From \eqref{EQ:hyper_closure}, with the solutions \eqref{EQ:vector_bps}, we find that the BPS configurations of the scalars are such that
\begin{align}
  \partial_\varphi q^a &= 0 \;, \cr
  \partial_\varphi\tilde{q}^a &= 0 \;.
\end{align}

With this ansatz for $q^a$, one can then write down the full BPS equations $\delta\psi_{\dot{a}} = \delta\tilde{\psi}_{\dot{a}} = 0$ and adopt, for the background gauge multiplet, the BPS locus \eqref{EQ:vector_background_sols}. By imposing the reality conditions \eqref{EQ:hyper_reality_conditions} one can eliminate the dependency from the auxiliary fields and find
\begin{align}\label{EQ:BPS_equations_hyper}
  (\partial_t - \mr{i}\zeta)q^+ + \beta\csc\theta\,\partial_\theta q^- &= 0 \;, \cr
  (\partial_t - \mr{i}\zeta^*)q^- - \beta\csc\theta\,\partial_\theta q^+ &= 0 \;, \cr
  (\partial_t + \mr{i}\zeta)\tilde{q}^+ + \beta\csc\theta\,\partial_\theta\tilde{q}^- &= 0 \;, \cr
  (\partial_t + \mr{i}\zeta^*)\tilde{q}^- - \beta\csc\theta\,\partial_\theta\tilde{q}^+ &= 0 \;,
\end{align}
where $\zeta$ is the $\mf{g}$-valued constant defined in \eqref{EQ:zeta_definition}. Without loss of generality, we can take $\mc{R}(\zeta)$ to be diagonal.

The equations above have solutions\footnote{To keep a clean notation, we avoid introducing additional indices. The fact that $\mc{R}(\zeta)$ is diagonal makes the interpretation of \eqref{EQ:BPS_sols_hyper} particularly simple: for each component of $q^\pm$ and $\tilde{q}^\pm$ (and consequently of their Fourier coefficients $u_{\pm,k}$ and $u_{\pm,k}$) one should select the appropriate diagonal element of $\mc{R}(\zeta)$.}
\begin{align}\label{EQ:BPS_sols_hyper}
  q^+(\theta,t) &= \sum_{k\in\mb{Z}} e^{-\mr{i}kt} \big[{+}u_{+,k}\cosh(\omega_k\cos\theta) + \mr{i} v_{+,k}\sinh(\omega_k\cos\theta)\big] \;, \cr
  q^-(\theta,t) &= \sum_{k\in\mb{Z}} e^{-\mr{i}kt} \big[{+}u_{-,k}\cosh(\omega_k\cos\theta) - \mr{i} v_{-,k}\sinh(\omega_k\cos\theta)\big] \;, \cr
  \tilde{q}^+(\theta,t) &= \sum_{k\in\mb{Z}} e^{+\mr{i}kt} \big[{+}u^*_{-,k}\cosh(\omega_{k}\cos\theta) + \mr{i} v^*_{-,k}\sinh(\omega_{k}\cos\theta)\big] \;, \cr
  \tilde{q}^-(\theta,t) &= \sum_{k\in\mb{Z}} e^{+\mr{i}kt} \big[{-}u^*_{+,k}\cosh(\omega_{k}\cos\theta) + \mr{i} v^*_{+,k}\sinh(\omega_{k}\cos\theta)\big] \;,
\end{align}
for
\begin{align}
  \omega_k = \beta^{-1}|k+\zeta|
\end{align}
and constants $u$'s and $v$'s, such that
\begin{align}\label{EQ:BPS_hyper_Fourier_coeffs}
  u_{-,k} &= \f{k+\zeta}{|k+\zeta|}\, v_{+,k} \;, \cr 
  v_{-,k} &= \f{k+\zeta}{|k+\zeta|}\, u_{+,k} \;. 
\end{align}

\section{Hypermultiplet Partition Function}\label{APP:3d_partition_function}
As a check of the localization argument that leads to the one-dimensional theory \eqref{EQ:1d_path_integral}, we perform a direct computation of the partition function for the action in \eqref{EQ:on_shell_action}. Our goal is to show that \eqref{EQ:on_shell_action} generates the same partition function $Z_{1\mr{d}}(\zeta,\zeta^*)$ as in \eqref{EQ:1d_partiton_function}, in agreement with our claim that the contribution of the fluctuations around the localization locus amounts to an overall constant.

We start by rewriting the on-shell lagrangian \eqref{EQ:on_shell_lagrangian} as
\begin{align}
  \mh{L}_{\mr{on}} = \tilde{q}^a\mc{B}_{ac}q^c + \tilde{\psi}^{\dot{a}}\mc{F}_{\dot{a}\dot{c}}\psi^{\dot{c}} \;,
\end{align}
in terms of the differential operators
\begin{align}
  \mc{B}_{ac} &= -\epsilon_{ac}\f{1}{r^2}\bigg(\triangle_{S^2} + \f{\mf{d}^2}{\beta^2} - \sigma^2\bigg) + \epsilon_{au}(\gamma^{\ms{3}})^u\vphantom{\epsilon}_c\f{\mf{d}}{r^2\beta} - \delta_{ac}\f{\sigma}{r^2} \;, \label{EQ:diff_op_B} \\
  \mc{F}_{\dot{a}\dot{c}} &= -\epsilon_{\dot{a}\dot{c}}\f{\mr{i}}{r} \bigg(\slashed{\mr{D}}_{S^2} + \gamma^{\ms{3}}\f{\mf{d}}{\beta}\bigg) + \epsilon_{\dot{a}\dot{u}}(\gamma^{\ms{3}})^{\dot{u}}\vphantom{\epsilon}_{\dot{c}}\f{\mr{i}\sigma}{r} \;, \label{EQ:diff_op_F}
\end{align}
which depend on the background vector multiplet components \eqref{EQ:vector_background_sols}.
We have introduced the shorthand
\begin{align}
  \mf{d} = \partial_t-\mr{i}a \;,
\end{align}
while with $\triangle_{S^2}$ and $\slashed{\mr{D}}_{S^2}$ we denote, respectively, the Laplace and the Dirac operator on the unit 2-sphere.

To determine the spectrum of \eqref{EQ:diff_op_B} and \eqref{EQ:diff_op_F}, one can decompose scalar and spinor fields in terms of orthogonal bases formed by Fourier modes along the $S^1$ and (spin-weighted) spherical harmonics on $S^2$. On such a base, the action of $\mf{d}$, $\triangle_{S^2}$ and $\slashed{\mr{D}}_{S^2}$ is trivial.
In particular, we remind the reader that a spherical harmonic of spin $s$,
\begin{align}
  Y^{(s)}_{\ell,m}(\theta,\varphi) = \;&(-1)^m \sqrt{\f{(\ell+m)!(\ell-m)!(2\ell+1)!}{4\pi(\ell+s)!(\ell-s)!}}\,\sin^{2\ell}(\theta/2) \cr
  &\times\sum_{j=0}^{\ell-s}\begin{pmatrix}\ell-s\\j\end{pmatrix}\begin{pmatrix}\ell+s\\j+s-m\end{pmatrix}(-1)^{\ell-j+s}e^{\mr{i}m\varphi}\cot^{2j+s-m}(\theta/2) \;,
\end{align}
is defined for $s$, $\ell$, $m$ half-integers, with $s-\ell$ and $s-m$ integers, and with the constraints $\ell\geq|s|$ and $|m|\geq\ell$.
Both $\triangle_{S^2}$ and $\slashed{\mr{D}}_{S^2}$ can be written in terms of raising and lowering operators $\eth$ and $\bar{\eth}$. These act on a function $f^{(s)}$ of definite spin $s$ with
\begin{align}
  \eth f^{(s)} &= -(\sin\theta)^{+s}\left(\f{\partial}{\partial\theta} + \f{\mr{i}}{\sin\theta}\f{\partial}{\partial\varphi}\right)[(\sin\theta)^{-s}\,f^{(s)}] \;, \cr
  \bar{\eth} f^{(s)} &= -(\sin\theta)^{-s}\left(\f{\partial}{\partial\theta} - \f{\mr{i}}{\sin\theta}\f{\partial}{\partial\varphi}\right)[(\sin\theta)^{+s}\,f^{(s)}] \;.
\end{align}
In particular,
\begin{align}
  \eth \, Y^{(s)}_{\ell,m}(\theta,\varphi) &= +\sqrt{(\ell-s)(\ell+s+1)} \; Y^{(s+1)}_{\ell,m}(\theta,\varphi) \;, \cr
  \bar{\eth} \, Y^{(s)}_{\ell,m}(\theta,\varphi) &= -\sqrt{(\ell+s)(\ell-s+1)} \; Y^{(s-1)}_{\ell,m}(\theta,\varphi) \;.
\end{align}

The partition function for the hypermultiplet can be expressed as
\begin{align}
  Z_{\mr{hyper}} = Z_{q} \, Z_{\psi}
\end{align}
in terms of the partition functions $Z_q$ and $Z_\psi$ associated, respectively, with the bosonic and the fermionic degrees of freedom in the (on-shell) multiplet.
These come from taking the functional determinant of the differential operators \eqref{EQ:diff_op_B} and \eqref{EQ:diff_op_F}.
By taking the product of the eigenvalues of \eqref{EQ:diff_op_B} and \eqref{EQ:diff_op_F}, counted with their multiplicity in $m$, we find
\begin{align}
  Z_{q} = \prod_{\rho\in\mc{R}}\,\prod_{k\in\mb{Z}}\;\prod_{\ell\in\mb{N}}\;&\left[\bigg(\f{(k+\rho(a))^2}{r^2\beta^2}+\f{\ell^2}{r^2}+\f{\rho(\sigma)^2}{r^2}\bigg)\bigg(\f{(k+\rho(a))^2}{r^2\beta^2}+\f{(\ell+1)^2}{r^2}+\f{\rho(\sigma)^2}{r^2}\bigg)\right]^{-2\ell-1} 
\end{align}
and
\begin{align}
  Z_\psi = \prod_{\rho\in\mc{R}\vphantom{\f{1}{2}}}\;\;\prod_{k\in\mb{Z}\vphantom{\f{1}{2}}}\,\prod_{\ell\in\mb{N}+\f{1}{2}}\;\left[\f{(k+\rho(a))^2}{r^2\beta^2}+\f{(\ell+\f{1}{2})^2}{r^2}+\f{\rho(\sigma)^2}{r^2}\right]^{2(2\ell+1)} \;.
\end{align}
In the above, we set $r=1$ to render the partition functions adimensional.

When combining the two, we obtain
\begin{align}
  Z_{\mr{hyper}} &= \prod_{\rho\in\mc{R}}\,\prod_{k\in\mb{Z}}\,\f{\beta^2}{|k+\rho(\zeta)|^2} \;.
\end{align}
The product over $k$ can be regularized as in \eqref{EQ:zeta_regularized_product}. The result,
\begin{align}
  Z_{\mr{hyper}} &= \prod_{\rho\in\mc{R}} \f{1}{4|\sin(\pi \rho(\zeta))|^2} \;,
\end{align}
agrees with the partition function \eqref{EQ:1d_partiton_function_full} obtained from the one-dimensional theory.

\bibliographystyle{utphys2}
\bibliography{main}

\end{document}